\documentclass[aps,pra, superscriptaddress, twocolumn]{revtex4-1}
\usepackage{ORI_Group_style}

\newcommand{\Phop}{\hat{\Phi}}
\newcommand{\Psop}{\hat{\Psi}}
\newcommand{\Psdop}{\hat{\Psi}^\dag}

\newcommand{\mmop}{\hat{\mmu}}
\newcommand{\muop}{\hat{\mu}}

\newcommand{\fop}{\hat{f}}
\newcommand{\fdop}{\fop^\dag}

\newcommand{\sigop}{\hat{\boldsymbol{\sigma}}}
\newcommand{\sigz}{\hat{\sigma}^z}
\newcommand{\sigm}{\hat{\sigma}^-}
\newcommand{\sigp}{\hat{\sigma}^+}

\newcommand{\U}{\hat{U}}

\newcommand{\rhop}{\hat{\rho}}

\newcommand{\wL}{\w_L}

\newcommand{\wsig}{\w_\sigma}
\newcommand{\wsigj}{\w_{\sigma j}}
\newcommand{\wtsig}{\tilde{\w}_\sigma}

\newcommand{\wj}{\w_j}

\newcommand{\keff}{\kappa_\text{eff}}
\newcommand{\geff}{g_\text{eff}}
\newcommand{\Weff}{\W_\text{eff}}

\newcommand{\mum}{\mu\text{m}}

\newcommand{\gr}{\gamma_0}
\newcommand{\gq}{\gamma_q}
\newcommand{\mmu}{\boldsymbol{\mu}}
\newcommand{\kp}{\kappa}

\newcommand{\bj}{\bold{j}}
\newcommand{\bdlt}{\boldsymbol{\delta}}
\newcommand{\bbeta}{\boldsymbol{\beta}}
\newcommand{\balp}{\boldsymbol{\alpha}}

\newcommand{\vphi}{\varphi}
\newcommand{\ve}{\varepsilon}

\begin{document}

\title{Hybrid Architecture for Engineering Magnonic Quantum Networks}

\author{C.~C.~Rusconi}
\affiliation{Institute for Quantum Optics and Quantum Information of the
Austrian Academy of Sciences, A-6020 Innsbruck, Austria.}
\affiliation{Institute for Theoretical Physics, University of Innsbruck, A-6020 Innsbruck, Austria.}
\affiliation{Max-Planck-Institut f\"ur Quantenoptik, Hans-Kopfermann-Strasse 1, 85748 Garching, Germany.}
\author{M.~J.~A.~Schuetz}
\affiliation{Physics Department, Harvard University, Cambridge, MA 02318, USA.}
\author{J.~Gieseler}
\affiliation{Physics Department, Harvard University, Cambridge, MA 02318, USA.}
\author{M.~D.~Lukin}
\affiliation{Physics Department, Harvard University, Cambridge, MA 02318, USA.}
\author{O.~Romero-Isart}
\affiliation{Institute for Quantum Optics and Quantum Information of the
Austrian Academy of Sciences, A-6020 Innsbruck, Austria.}
\affiliation{Institute for Theoretical Physics, University of Innsbruck, A-6020 Innsbruck, Austria.}

\date{\today}

\begin{abstract} 
We theoretically show that a network of superconducting loops and magnetic particles can be used to implement magnonic crystals with tunable magnonic band structures. 
In our approach, the loops mediate interactions between the particles and allow magnetic excitations to tunnel over long distances.
 As a result, different arrangements of loops and particles allow one to engineer the band structure for the magnonic excitations.
Furthermore, we show how magnons in such crystals can serve as a quantum bus for long-distance magnetic coupling of spin qubits. 
The qubits are coupled to the magnets in the network by their local magnetic-dipole interaction and provide an integrated way to measure the state of the magnonic quantum network. 
\end{abstract}

\maketitle

\section{Introduction}

Complex microscopic interactions between particles inside materials often give rise to emergent collective excitations. 
This collective behavior can be effectively described in terms of weakly interacting quasi-particles which propagate freely in the surrounding medium and follow dispersion relations which are determined by the microscopic details~\citep{Kittel1987}. This treatment allows to greatly simplify the description of otherwise intractable problems~\citep{Fetter2012}.
In many cases, the dispersion relation of quasi-particles can be tailored by a careful design of the host medium. 
For example, photonic crystals~\citep{Joannopoulos2011} are engineered materials where the propagation of photons is artificially designed by periodically arranging materials with different refractive indices~\citep{Chang2018}.
Quantum emitters can then be coupled to such structures for a variety of applications ranging from quantum simulation~\citep{Douglas2015,Hood2016,Hung2016,Schuetz2017}  and quantum information processing~\citep{Schuetz2015,Yao2012,Manenti2017} to the study of open quantum system~\citep{GonzalezTudela2017PRA,Sajeev1994}.

Magnons, collective excitations of magnetization in magnetically ordered materials, have recently attracted significant attention in the context of quantum information science.
Strong quantum coherent coupling of magnons to a microwave resonator~\citep{Huebl2013,Tabuchi2014,Lambert2016,Zhang2016,Zhang2015,Rameshti2015,Soykal2010}, optical photons~\citep{Kusminskiy2016,Osada2016,Haigh2018}, and superconducting qubits~\citep{Tabuchi2015,Tabuchi2016}, have been recently reported. 
Magnonic systems~\citep{Nikitov2001,Krawczyk2014,Chumak2017} with tailored magnonic propagation properties are also investigated as a magnon quantum bus to couple quantum emitters over long distances~\citep{Andrich2017,Kosen2018,Trifunovic2013}.
In present magnonic systems, spin wave propagation between the ferromagnetic elements is mediated by dipolar coupling.
Thus, sufficiently high coupling over long distances requires ferromagnets with high saturation magnetization. However, those materials suffer from high losses~\cite{Qin2018,Topp2010,Gubbiotti2007}. In contrast, materials such as YIG have little loss but also have a small saturation magnetization and thus a lower magnetic dipole coupling.

\begin{figure}[b]
	\includegraphics[width=\columnwidth]{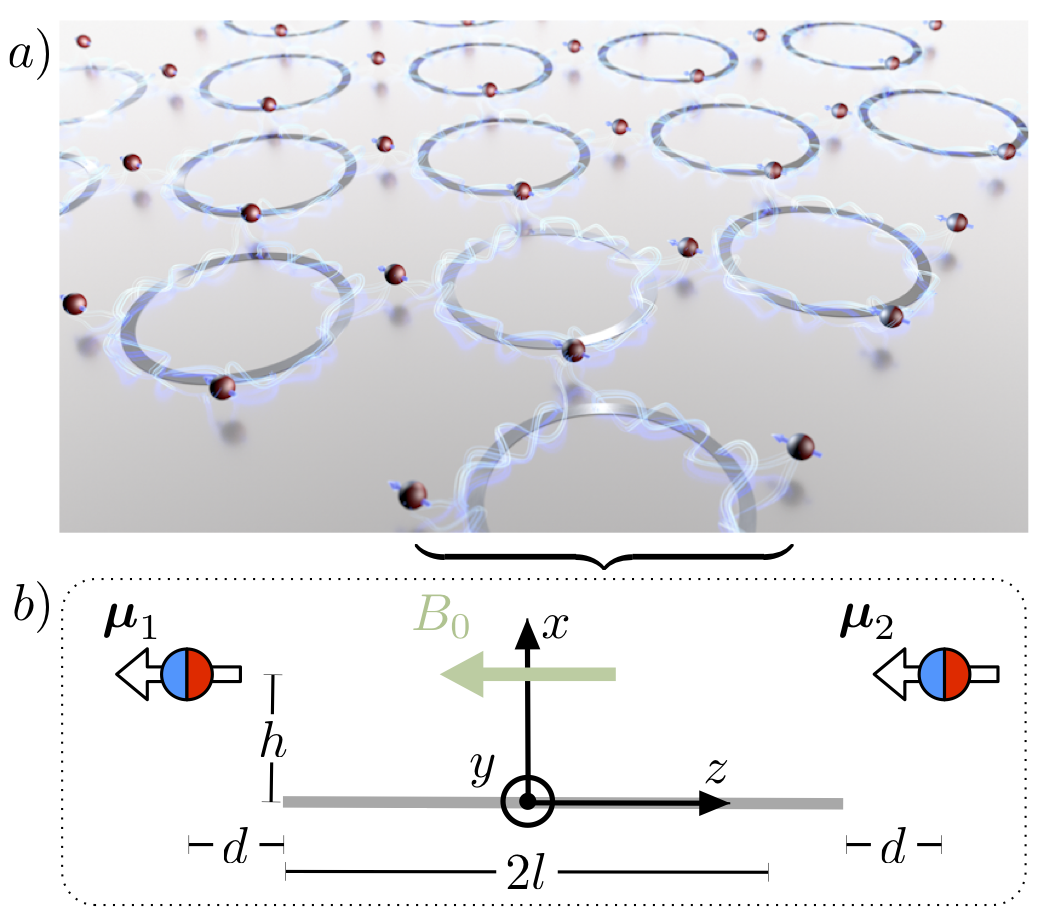}
	\caption{a) Schematic illustration of a general hybrid magnetic lattice (HML) of superconducting-loops and magnetic particles. b) Scheme of the simplest cell of the HML: two magnetic particles positioned at a distance $d$ and height $h$ from two opposite points of a superconducting ring of radius $l$.}
	\label{Fig:Illustration} 
\end{figure}

In this article, we propose a network of superconducting loops~\citep{Davidovic1996} that couples magnetic particles over distances larger than what can be achieved with magnetic dipole-dipole interactions in free space.
This allows to combine low loss materials such as YIG with the desired long range coupling.
In such a set-up, the excitation of the collective magnetization in a particle tunnels to other particles provided that there is a superconducting loop between them.
This provides a lot of flexibility in the topology of the networks that can be realized with this architecture thereby enabling a wide range of applications.
First we describe how to engineer artificial magnonic crystals using a periodic arrangement of magnetic particles and superconducting loops called hereafter hybrid magnetic lattice (HML), as shown in \figref{Fig:Illustration}.a. 
Second, we discuss how to interface spin qubits with HMLs via their dipolar coupling to the magnetic particles. The coupling enables long-range magnetic coupling of spin qubits and it introduces quantum non-linear elements into the magnonic crystal. In this context, our proposal offers an \textit{all-magnetic} solid-state alternative to optical quantum emitters coupled to photonic crystals~\citep{Chang2018}. In addition, the tunability of the magnonic band-gap by an externally applied magnetic field offers a handle which has no analogue in photonic systems. 
Through this feature, the qubit frequency can be tuned to lie inside or outside the band-gap, making the qubit dynamics predominantly conservative or dissipative, respectively~\citep{Chang2018}. Importantly, the band-gap can be tuned \textit{on-demand} in real time, thereby giving direct access to various, very different many-body problems simply by varying the external magnetic bias field.

The article is structured as follows. In \secref{sec:Model}, we discuss how magnonic crystals with a tailored band structure can be designed from a hybrid lattice of superconducting loops and magnetic particles. We first introduce the Hamiltonian describing tunneling of magnons between different magnetic particles via a loop-mediated interaction. Then, we present some specific examples of HMLs. In~\secref{sec:Qubits}, we discuss in detail the coupling between a qubit and a magnetic particle and show how to couple two distant spin qubits via the HML.
Finally, we draw our conclusions in \secref{sec:Conclusions}. Further details are provided in the appendices. 

\section{Artificial Magnonic Crystal}\label{sec:Model}

In this section, we focus on how to engineer an HML by placing magnetic particles near superconducting loops. In~\secref{SSec:Hamiltonian}, we present the Hamiltonian describing the interaction between a single superconducting loop and several magnetic particles. In~\secref{SSec:Examples}, we generalized the Hamiltonian to many loops and use this model to discuss specific HML examples.

\subsection{Hamiltonian of the elementary cell: one superconducting loop and several magnets} \label{SSec:Hamiltonian}

We consider $N$ magnetic particles with magnetic moments $\mmu_j$ ($j=1,\ldots,N$) located at positions $\rr_j$ near a superconducting circular loop with radius $l$. An external bias field $\BB_0\equiv-B_0\uez$ is applied parallel to the plane containing the loop (see the illustration for $N=2$ in~\figref{Fig:Illustration}.b).
We model the superconductive ring as a single mode $LC$-resonator, whose self-inductance $L$ and capacitance $C$ are of geometrical origin (see~\appref{apdx:ModelLoop}). 
We model the magnetic particle as a sphere of radius $R$, whose center of mass position $\rr_j$ lies outside the area encircled by the loop. The applied field $\BB_0$ polarizes the magnetic particles uniformly. Therefore, the magnetic field produced by the particles can be approximated as the field generated by a constant magnetic point dipole of magnitude $\mu_j=\vert\mmu_j\vert$.
The coherent dynamics of the system is modeled by the following quantum mechanical Hamiltonian (see~\appref{apdx:Ham_Derivation} for a derivation)
\be\label{eq:Hamiltonian}
\begin{split}
	\Hop =& \frac{\Qop^2}{2C} + \frac{1}{2L}\Big[\Phop - \sum_j \Phi_j(\mmop_j)\Big]^2-\sum_j \mmop_j \cdot \BB_0\\
	&+\frac{1}{2L}\bigg[\sum_{j=1}^N \Phi_j(\mmop_j)\bigg]^2\!+U_\text{d}(\{\mmop_j\})+\!\sum_j \Vop^j_a(\mmop_j).
\end{split}
\ee
Here, $\Qop$ ($\Phop$) is the charge (flux) operator of the loop~\citep{Vool2017}, and
\be\label{eq:Phi_j}
	\Phi_j(\mmop_j) \equiv \Phi^j_\text{bias}\id+\delta\Phi_j(\Delta\mmop_j)
\ee
is the external flux induced in the coil by the magnetic dipole moment $\mmop_j$. In \eqnref{eq:Phi_j}, $\Phi^j_\text{bias}$ is the field induced in the loop by the magnetic moment at its  equilibrium value $\avg{\mmop_j}_0$, while $\delta\Phi_j(\Delta\mmop_j)$ is the flux induced in the loop by the magnetic moment fluctuation $\Delta \mmop_j\equiv \mmop_j -\avg{\mmop_j}_0$.
The first two terms in \eqnref{eq:Hamiltonian} represent the energy of the loop in the presence of the magnets. 
The third term refers to the Larmor precession of the magnetic moments about the direction of $\BB_0$. 
The fourth term $[\sum_j \Phi_j(\mmop_j)]^2/2L$ represents the total loop-mediated magnetic interaction between the magnets. The magnetic dipole interaction between the magnets, obtained after tracing out all the electromagnetic field modes, is modified (as compared to free space) due to the presence of the loop, which is treated as a single electromagnetic field mode (see~\appref{apdx:ModelLoop}). The correction to the free dipole-dipole interaction is precisely the fourth term in \eqnref{eq:Hamiltonian}. The seemingly additional dipole-dipole interaction included in the second term of \eqnref{eq:Hamiltonian} is cancelled out perfectly when tracing out the loop's electromagnetic field mode (see~\appref{apdx:Ham_Derivation} for a detailed derivation). This subtle point was previously discussed in the literature in \citep{DeBernardis2018}.
The remaining contribution to the  magnetic dipole-dipole interaction is mediated by the free space electromagnetic modes and is represented by $U_\text{d}(\{\mmop_j\})$ in \eqnref{eq:Hamiltonian}.
The last term in \eqnref{eq:Hamiltonian} is the magnetic anisotropy energy of each particle which represents the energy cost of magnetizing the particle along a certain direction due to the interaction between its magnetic moment and its internal crystal structure~\citep{Chikazumi}.

Let us now introduce the macrospin $\FFop$ of a magnetic particle, which is related to the magnetic moment by the gyromagnetic relation $\mmop_j = \hbar \gr \FFop_j$~\footnote{We define the spin operator to be dimensionless, namely $[\Fop_i,\Fop_j]=\im \levi{ijk}\Fop_k$.}. In the following, we assume the magnetic particles to be identical, namely they have the same gyromagnetic ratio $\gr$, the same radius $R$, and thus the same total spin $\FFop^2=F(F+1)\id$, where we define $F\equiv \mu/(\hbar \gr)$ and $\mu_j\equiv \mu$ $\forall j$. The flux fluctuations in \eqnref{eq:Phi_j} can be written as $\delta\Phi_j(\Delta\mmop_j) = \Phi_{\text{e}j} \sum_\nu I_j^\nu\Delta\Fop_j^\nu$ ($\nu=x,y,z$), where $\Delta \FFop_j \equiv \Delta\mmop_j/(\hbar \gr)$, $\Phi_{\text{e}j}\equiv \hbar \gr \mu_0/ 4\pi d_j$. Here, $d_j$ is the smallest distance, in the plane containing the loop, between the dipole's position and a point in the loop (see~\figref{Fig:Illustration}.b). $I_j^\nu$ is a dimensionless geometrical factor which contains the dependence on the center of mass position of the nanomagnet and on the orientation of its magnetic moment (see~\appref{apdx:FluxCoil}).

For a sufficiently large $B_0$, such that the thermal energy is negligible compared to the interaction $-\mmop_j\cdot\BB_0$, the fluctuations of the magnetic moment $\Delta \mmop_j$ can be expressed within the Holstein-Primakoff approximation as $\Delta\hat{\mu}_j^z = \hbar \gr \fdop_j\fop_j$, $\Delta \hat{\mu}_j^x = \hbar \gr\sqrt{2F} (\fdop_j+\fop_j)/2$, and $\Delta \hat{\mu}_j^y = \hbar \gr \sqrt{2F}(\fdop_j-\fop_j)/(2\im)$. The operator $\fop_j$ ($\fdop_j$) creates (annihilates) an excitation in the uniformly precessing (Kittel) magnonic mode of the $j$-th magnet, and satisfies $[\fop_i,\fdop_j]=\delta_{ij}$.
Within the Holstein-Primakoff approximation (valid when $\avg{\fdop_j\fop_j}\ll 2F$) and the assumption that the $LC$-circuit is far detuned from the magnonic modes (such that the degrees of freedom of the circuit can be traced out), the coherent dynamics of the magnets reduce to
 \be 
\Hop_\text{M} = \hbar\sum_{j=1}^N \wj \fdop_j\fop_j + \hbar \sum_{j\neq i=1}^N\pare{\mathcal{J}_{ij}+\mathcal{J}_{ij}^\text{d}}\fdop_i\fop_j +\Vop_\text{lin}.
 \ee
Counter-rotating terms (of the form $\fdop_i\fdop_j+\fop_i\fop_j$) have been neglected within the rotating-wave approximation (see \appref{apdx:Hamiltonian_SLNM}). Here, $\Vop_\text{lin}$ is a linear term in the bosonic operators which can be reabsorbed by defining a new equilibrium position $\avg{\FFop}_0$. Notably, for the particular case of the magnetic particles lying in the plane containing the loops one finds that $\Phi_\text{bias}^j\!=\!0\,\forall j$ and $\Vop_\text{lin}$ disappears (see~\appref{apdx:Hamiltonian_SLNM}). It is thus always possible to write the quadratic Hamiltonian describing the magnon dynamics in a HML as  
\be\label{eq:QuadHam}
\Hop_\text{M} = \hbar\sum_{j=1}^N \wj \fdop_j\fop_j + \hbar \sum_{j\neq i=1}^N\pare{\mathcal{J}_{ij}+\mathcal{J}_{ij}^\text{d}}\fdop_i\fop_j.
\ee
Here, $\wj$ is the sum of the frequency associated with the magnetic anisotropy and the Larmor precession frequency due to the total magnetic field, which consists of the external field $\BB_0$, the field created by other magnets and the field created by the superconducting loop (see~\appref{apdx:Hamiltonian_SLNM}). The second term in \eqnref{eq:QuadHam} describes magnon tunneling between magnets. The total tunneling rate has two contributions. The contribution from the free space magnetic dipole-dipole interaction is given by $\mathcal{J}_{ij}^\text{d}\equiv-\hbar \gr^2\mu_0F(3\sin^2\theta_{ij}-2)/(8\pi r_{ij}^3)$, where $r_{ij}\equiv \vert\rr_i-\rr_j\vert$ and $\theta_{ij}$ is the angle between $\rr_i-\rr_j$ and $\uez$.
The contribution from the loop-mediated magnetic interaction is given by (see~\appref{apdx:Hamiltonian_SLNM})
\be\label{eq:J}
    \mathcal{J}_{ij} \equiv \pare{\frac{\hbar \gr \mu_0}{4\pi}}^2\frac{I_{ij}}{2\hbar d_i d_j L}F.   
\ee
Here, $I_{ij} \equiv I_i^*I_j $, where $I_j\equiv I_j^x+\im I_j^y$, depends on the mutual position of the magnetic particles at the sites $i,j$ and on the orientation of their magnetic moments (see~\appref{apdx:FluxCoil}). In particular, $\mathcal{J}_{ij}$ can be made independent of $i,j$ for symmetric arrangements of magnetic particles around the loop such that $I_{ij}\equiv I$ and $d_j=d$ $\forall j$ (see~\secref{SSec:Examples}).
We stress that $\mathcal{J}_{ij}$ scales as $1/(d_i d_j l)$, where the factor $1/(d_id_j)$ arises from the $1/d_j$-dependence of $\Phi_j(\mmop_j)$ and the factor $1/l$ arises from the linear dependence of $L$ on the loop radius [see~\eqnref{eq:L}]. For fixed $d_i,d_j\ll l$, the loop-mediated interaction thus leads to a magnon tunneling rate which scales as $\sim 1/r_{ij}$. The minimal possible distance $d$ (and thus the maximum achievable tunneling rate for a given loop geometry) is ultimately set by the critical field tolerated by the loop's wire in the Meissner state (see~\appref{apdx:critical_distance}).

As an example, let us consider the simple configuration of one loop and two magnetic particles illustrated in \figref{Fig:Illustration}.b for the particular case of $h=0$. For this case one has, $\wj = 2 \gr k_a/M_\text{s} + \gr B_0 + \mathcal{J}_{jj} - \hbar \gr^2\mu_0F/[16\pi(l+d)^3]$, $r_{12} = 2(l+d)$, $\theta_{12}=0$, and $\mathcal{J}_{12} = [\hbar \gr \mu_0/(4\pi d)]^2 I F /(2\hbar L)$, where $I_{12}=I_{21}\equiv I$. 
The geometrical inductance of a circular coil is approximated as $L \approx \mu_0 l \ln\pare{8l/\tau} $ for $\tau \ll l$, where $\tau$ is the wire thickness. For $l\gg d$, one finds
\be
	\frac{\mathcal{J}_{12}}{\mathcal{J}^\text{d}_{12}} \approx \frac{(l/d)^2 2 I^2}{\pi \ln(8l/\tau)} \gg 1.
\ee
For $R=1\mum$, $d=1.5\mum$, $l=30\mum$, $h=0$, and $\tau=50\text{nm}$, which leads to $I\approx 1.9$~\footnote{With this configuration the field intensity at the surface of the coil generated by the magnet is $B \approx 50 \text{mT}$~\appref{apdx:critical_distance}. Depending on the distance between the particle and the loop might be necessary to consider the loop to be made of a high-Tc superconductor (see~\appref{apdx:critical_distance}).}, the tunneling rate due to the inductive magnetic interaction is then $\mathcal{J}_{12}/2\pi\approx 5.85 ~\text{MHz}$ whereas the one due to the magnetic dipole interaction is $\mathcal{J}^d_{12}/2\pi\approx 0.09 ~\text{MHz}$. In general, for sufficiently large loop size, magnetic dipole-dipole interactions are negligible as compared to loop mediated coupling (see~\figref{Fig:Bone}.b in \appref{apdx:Bone}). In this case, the magnon tunneling can be approximated by $\mathcal{J}_{ij}$. We remark that larger tunneling rates could be obtained by inscribing the magnets in the contour defined by the loop (see~\appref{apdx:Bone}). However, this configuration will not be considered further since it is not well  suited for building large networks.

\subsection{Examples of hybrid magnetic lattices}\label{SSec:Examples}

Let us now focus on how to build networks by periodic arrangements of superconducting loops and magnetic particles. \eqnref{eq:Hamiltonian} can be directly generalized to the case of many superconducting loops thus yielding the general Hamiltonian of a HML. In the following, we neglect the magnetic dipole-dipole coupling ($\mathcal{J}_{ij}^d=0$) and the flux generated in a coil by next-to-nearest neighbor magnets as well as by neighboring superconducting coils. Furthermore, within the assumption of identical loops, magnetic particles, and relative positioning of particles and loops, the magnon frequency (tunneling rate) is site-independent, namely $\w_j\equiv\w_0$ $\forall j$ ($\mathcal{J}_{ij}\equiv\mathcal{J}$ $\forall i,j$).

\begin{figure*}
	\includegraphics[width=2\columnwidth]{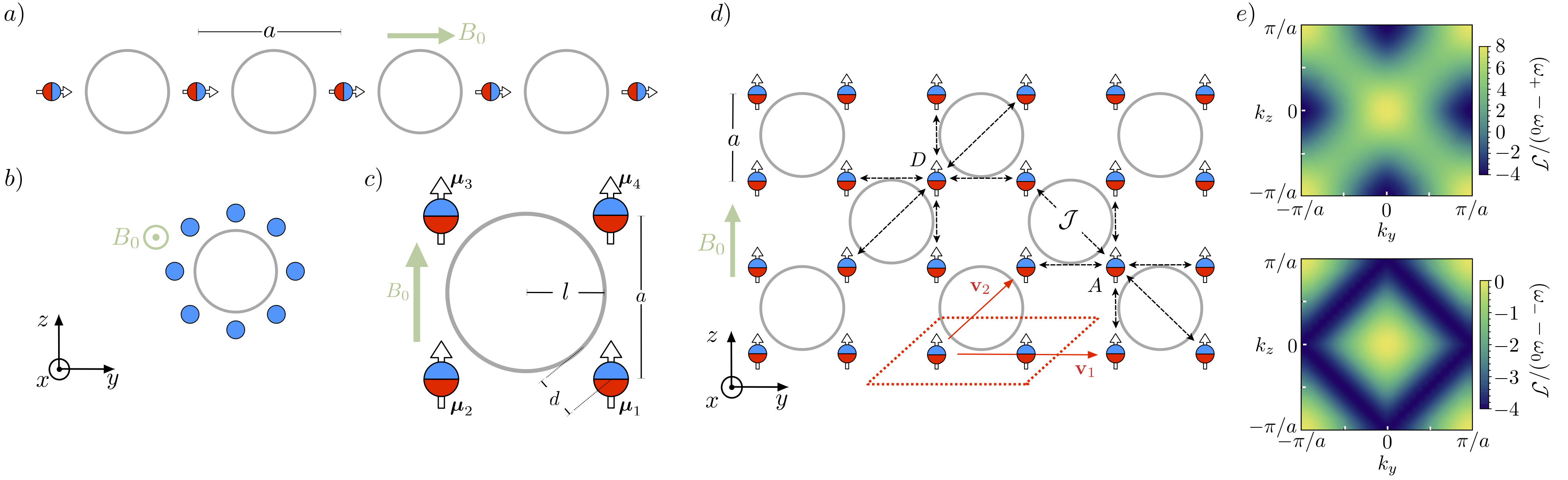}
	\caption{a) Top view of a 1D HML. b) Top view of a system of vertically polarized magnets around a common loop. c) Unit cell of a 2D HML comprising one loop and four magnetic particles. d) Top view of a 2D HML implementing a 2D magnonic crystal by a checkboard arrangement of the single cell in c). The black dashed arrows indicate coupling between nearest-neighbors at the same rate $\mathcal{J}$ and the red dashed area corresponds to the elementary cell of such a magnonic crystal with Bravais vectors $\bold{v}_1$ and $\bold{v}_2$. e) Plot for the magnonic bands $(\w_\pm-\w_0)/\mathcal{J}$ of the 2D HML in d).}
	\label{Fig:Schematic} 
\end{figure*}

In the following we consider three different examples of HMLs:

(i) A one dimensional HML, shown in \figref{Fig:Schematic}.a, can be described by 
\be 
\Hop_\text{M}^{1\text{D}} = \hbar \w_0 \sum_{j} \fdop_j\fop_j + \hbar \mathcal{J}\sum_{j} (\fdop_j\fop_{j+1}+\fdop_{j+1}\fop_j).
\ee
This textbook Hamiltonian describes magnon tunneling to nearest-neighbors in a one dimensional crystal with $N$ lattice sites separated by a distance $a=2(d+l)$. Assuming periodic boundary conditions, $\Hop_\text{M}^{1\text{D}}$ can be diagonalized in the reciprocal space leading to a magnon dispersion relation $\w(k) =\w_0 +2\mathcal{J}\cos(ka)$, where $k=2\pi n/(Na)$ ($n\in[N/2,N/2-1]$). 
In the continuum limit ($N\gg 1$), the magnon propagation is thus restricted to the frequency band $\w \in [\w_0-2\mathcal{J},\w_0+2\mathcal{J}]$, which can be tuned in real time by simply modifying the external magnetic field $B_0$, and hence $\w_0$. Note that this in-situ tunability is a characteristic feature of the proposed HMLs in this article.

(ii) A HML where $N$ magnets couple to each other with the same strength can be realized with the circular geometry shown in \figref{Fig:Schematic}.b. 
The Hamiltonian is given by 
\be 
\Hop_\text{M}^{N\text{D}} = \hbar \w_0 \sum_{j=1}^{N} \fdop_j\fop_j+ \hbar \mathcal{J}\sum_{i\neq j=1}^N\fdop_i\fop_j,
\ee 
with an all-to-all interaction $\sim \mathcal{J}$ [cf. \eqnref{eq:J}].
Here, the geometrical factor $I_{ij}=I$ is different from example (i) due to the fact that the magnets are now polarized perpendicularly to the superconducting coils. 
In principle, this Hamiltonian could be used to generate magnonic superradiance by enhancing dissipation in the coil and allowing the system to evolve beyond the quadratic approximation~\citep{Gross1982}.

(iii) A two-dimensional HML can be realized by a repetition of the single cell in~\figref{Fig:Schematic}.c leading to the structure displayed in~\figref{Fig:Schematic}.d.  Owing to the checkerboard arrangements of superconducting loops, we distinguish two magnonic sublattices: magnons in the $D$ ($A$) sublattice preferably tunnel along the direction of the main diagonal (anti-diagonal) in the $yz$-plane.
This HML can thus be described as a 2D Bravais lattice with a basis where each elementary cell contains the two types of sites $D$ and $A$ (see \figref{Fig:Schematic}.d). 
The operators $\fop^\text{A}_\bj,\fop^{\text{A}\dag}_\bj$ ($\fop^\text{D}_\bj,\fop^{\text{D}\dag}_\bj$) respectively create and annihilate a magnon in the sublattice $A$ ($D$) within the cell at position $\bj=j_y \bold{v}_1+j_z\bold{v}_2\equiv (j_y,j_z)$, ($j_y,j_z\in \mathbb{Z}$), where $\bold{v}_1 = (2a,0)$ and $\bold{v}_2=(a,a)$ are Bravais vectors.
The Hamiltonian of this 2D HML is given by (see~\appref{apdx:2D_SLNM})
\be\label{eq:Ham2D}
\begin{split}
    \Hop_\text{M}^{2\text{D}} =& \hbar \w_0 \sum_\bj\pare{ \fop^{\text{D}\dag}_\bj\fop^\text{D}_\bj + \fop^{\text{A}\dag}_\bj \fop^\text{A}_\bj }\\
    &+ \hbar\mathcal{J}\bigg[\sum_{\bj,\bbeta} \fop^{\text{D}\dag}_\bj \fop^\text{A}_{\bj+\bbeta}+\sum_{\bj,\balp} \fop^{\text{A}\dag}_\bj\fop^\text{A}_{\bj+\balp}\\
    &+\sum_{\bj,\bdlt} \fop^{\text{D}\dag}_\bj \fop^\text{D}_{\bj+\bdlt}+\hc \bigg].
\end{split}
\ee
Here, $\bbeta \in \{(\pm 1/2,\mp 1),(\pm 1/2,0)\}$, $\balp \in \{(\mp1,\pm 1)\}$, and $\bdlt \in \{(\pm1,0)\}$, 
with $\balp$ ($\bdlt$) and $\bbeta$ connecting the nearest neighbors of a point along the main anti-diagonal (diagonal) and along the $z$,$y$ direction in the basis specified by $\bold{v}_1$ and $\bold{v}_2$ (\figref{Fig:Schematic}.d). 
The magnon dispersion relation of \eqnref{eq:Ham2D} leads to two bands given by $\w_\pm(\kk)= \w_0+ 2\mathcal{J}[4\cos(k_x a)\cos(k_y a)\pm\sqrt{\Lambda}]$, where $\Lambda\equiv 4+4\cos(k_x a)\cos(k_y a)-\cos(2k_y a)-\cos(k_x a)+2\cos(2k_x a)\cos(2k_y a)$, with $a=\sqrt{2}(l+d)$. 
As shown in \figref{Fig:Schematic}.e, the upper band $\w_+(\kk)$ features saddle points at $\kk=(\pm\pi/2a,\pm\pi/2a)$ where the density of state diverges~\cite{Grosso2000}. 
As recently shown in~\citep{GonzalezTudela2017PRL}, this type of exceptional points may give rise to very exotic features in the quantum dynamics of emitters coupled to a two dimensional crystal.

\section{Spin qubits interfaced with a hybrid magnetic lattice}\label{sec:Qubits}

Our three examples show that HMLs can be engineered to realize artificial long-range magnonic crystals.
Let us now address how to magnetically interface spin qubits with the magnons in a given HML.
In \secref{sec:MagnonQubit_SiteCoupling}, we describe the local coupling between a spin qubit and a magnetic particle in a single site of a HML. In \secref{sec:Dissipation}, we discuss the sources of dissipation of the system. In \secref{sec:SWAP}, we analyse the magnon-mediated qubit-qubit interaction.

\subsection{Magnon-qubit coupling at a single site}\label{sec:MagnonQubit_SiteCoupling}

A spin qubit is coupled to a magnetic particle in a HML by local magnetic dipole-dipole interactions.
Specifically, we consider the interaction between the $j$-th magnet and an NV-center spin qubit, that is obtained from the $\{\ket{0},\ket{-1}\}$ subspace of the NV ground state triplet~\cite{Gruber2012}, placed at a position $\rr_q$ with respect to the center of the magnet.
The Hamiltonian of this system is given by
\be\label{eq:H_MQ_general}
\Hop_\text{MQ}^{(j)} =  \frac{\hbar}{2}\w_\text{q}\sigz_j-\frac{\hbar}{2} \gq \sigop_j\cdot \BB(\rr_q,\mmop_j),
\ee
where $\w_\text{q}\equiv \Delta_\text{NV}-\gq B_0$, $\gq$ is the qubit gyromagnetic ratio (generally different from $\gr$), $\Delta_\text{NV}$ the NV-center zero field splitting, and $\BB(\rr_q,\mmop_j)$ the magnetic field generated by the magnet at the position of the qubit.
Within the Holstein-Primakoff approximation, the rotating wave approximation, and assuming the qubit to be positioned along the x-axis of a reference frame centered in the magnet, $\rr_q=r_q\uex$, and oriented as in \figref{Fig:Illustration}.b (see~\appref{apdx:NV-Magnon} for the generalization to any other position), \eqnref{eq:H_MQ_general} is approximated by the Jaynes-Cumming Hamiltonian (see~\appref{apdx:NV-Magnon})
\be\label{eq:H_MQ}
\Hop_\text{MQ}^{(j)} = \inv{2}\hbar\wsigj \sigz_j - \hbar g (\fdop_j \sigm_j+\hc).
\ee
Here, $g \equiv 3\hbar \gr \gq \mu_0 \sqrt{2F}/(8\pi r_q^3)$ and the qubit frequency $\wsigj = \w_\text{q}+\hbar\gr\gq \mu_0F/(4\pi r_q^3)$ already contains the shift introduced by the dipole-interaction.
The dynamics of a general 2D HML with magnetically coupled spin qubits at each lattice site is described by the Jaynes-Cumming-Hubbard Hamiltonian $\Hop_\text{T} = \Hop_\text{M}+\sum_\bj \Hop_\text{MQ}^{(\bj)}$, namely in $k$-space [see \eqnref{eq:JCH} for the expression in real space]
\be\label{eq:JCH_kspace}
\begin{split}
	\Hop_\text{T} =& \hbar \sum_{\nu,\kk} \w_\nu(\kk) \fdop_{\nu\kk}\fop_{\nu\kk} +\hbar\sum_\bj \frac{\wsig}{2} \sigz_\bj\\
	& -\hbar \sum_{\nu,\bj,\kk}\pare{g_{\nu\bj\kk}\fdop_{\nu\kk} \sigm_\bj +\hc}.
\end{split}
\ee
Here, we introduced the $k$-space magnonic operator $\fop_{\nu\kk} = (1/N)\sum_\bj \fop_{\nu\bj} \exp(-\im a\bj\cdot\kk)$, which creates a magnon of momentum $\kk$ in the $\nu$-magnonic band propagating in a $N\times N$ 2D lattice characterized by the dispersion relation $\w_\nu(\kk)$, and the coupling rate $g_{\nu\bj\kk}\equiv(g_\nu/N)\exp(-\im a\bj\cdot\kk)$, where $g_\nu$ is the local coupling to a magnon in the $\nu$-band, $a$ is the HML lattice constant, and $\bj$ labels the sites in a 2D HML.
In \eqnref{eq:JCH_kspace}, we neglected the small interaction between the qubit and the loop as well as counter-rotating terms of the form $\sigp_\bj\fdop_{\nu\kk}+\sigm_\bj\fop_{\nu\kk}$  within the rotating wave approximation, valid provided $g,\vert\w_\nu(\kk)-\wsig\vert\ll \w_\nu(\kk)$. 

\subsection{Sources of dissipation}\label{sec:Dissipation}

The Hamiltonian \eqnref{eq:JCH_kspace} can lead to strongly correlated, coherent magnon physics \citep{Angelakis2007,Greentree2006}, provided that the relevant decoherence rates are sufficiently small compared to the coherent coupling rates of the system. 
While the coherent magnon tunneling $\mathcal{J}$ can reach several MHz (as discussed above), the coherent magnon-qubit coupling can be quantified as 
$g/(2\pi)\approx 5.2\times10^2(R[\text{nm}]^{1/2}/r_q[\text{nm}])^3 \text{MHz}$,
as a function of both the magnet size $R$ and magnet-qubit distance $r_q>R$; see caption of \figref{Fig:Scaling} for the remaining parameters. 
The main sources of decoherence arise from qubit dephasing and magnon decay, 
as any potential damping in the superconducting loop is suppressed by its large detuning.
For a NV-center spin qubit, characteristic dephasing times $T_2^*\approx 200\,\mu\text{s}$ have been reported~\citep{Zhao2012}, which can further be increased by dynamical decoupling schemes up to $T_2\approx 0.5\,\text{s}$~\citep{Abobeih2018}. 
In the low-temperature regime $\sim 1\mathrm{K}$ the magnon linewidth~\citep{Boventer2018} for a millimeter-size pure single-crystal YIG sphere has been measured as $\kp/2\pi \approx 0.5\,\text{MHz}$~\citep{Tabuchi2014}, 
at a relatively high magnon frequency of $\sim 10\text{GHz}$; 
this number could potentially be further reduced by working at lower frequencies according to the linear frequency dependence of the Gilbert damping rate in YIG~\citep{Klingler2017}. 
Accordingly, the regime $\mathcal{J}> \pi/T_2^*,\kp$ is within reach for particles of size $R\approx 1\mum$ (see~\secref{SSec:Hamiltonian}) with current experimental capabilities, 
while the regime $g > \kp$ is found to be challenging with the current reported values of the magnon linewidth.
However, the detrimental effects due to magnon decay can be reduced efficiently by operating in the dispersive regime, as detailed next.

\subsection{Effective qubit-qubit interaction thorugh a single cell of a HML}\label{sec:SWAP}

\begin{figure}
	\includegraphics[width=\columnwidth]{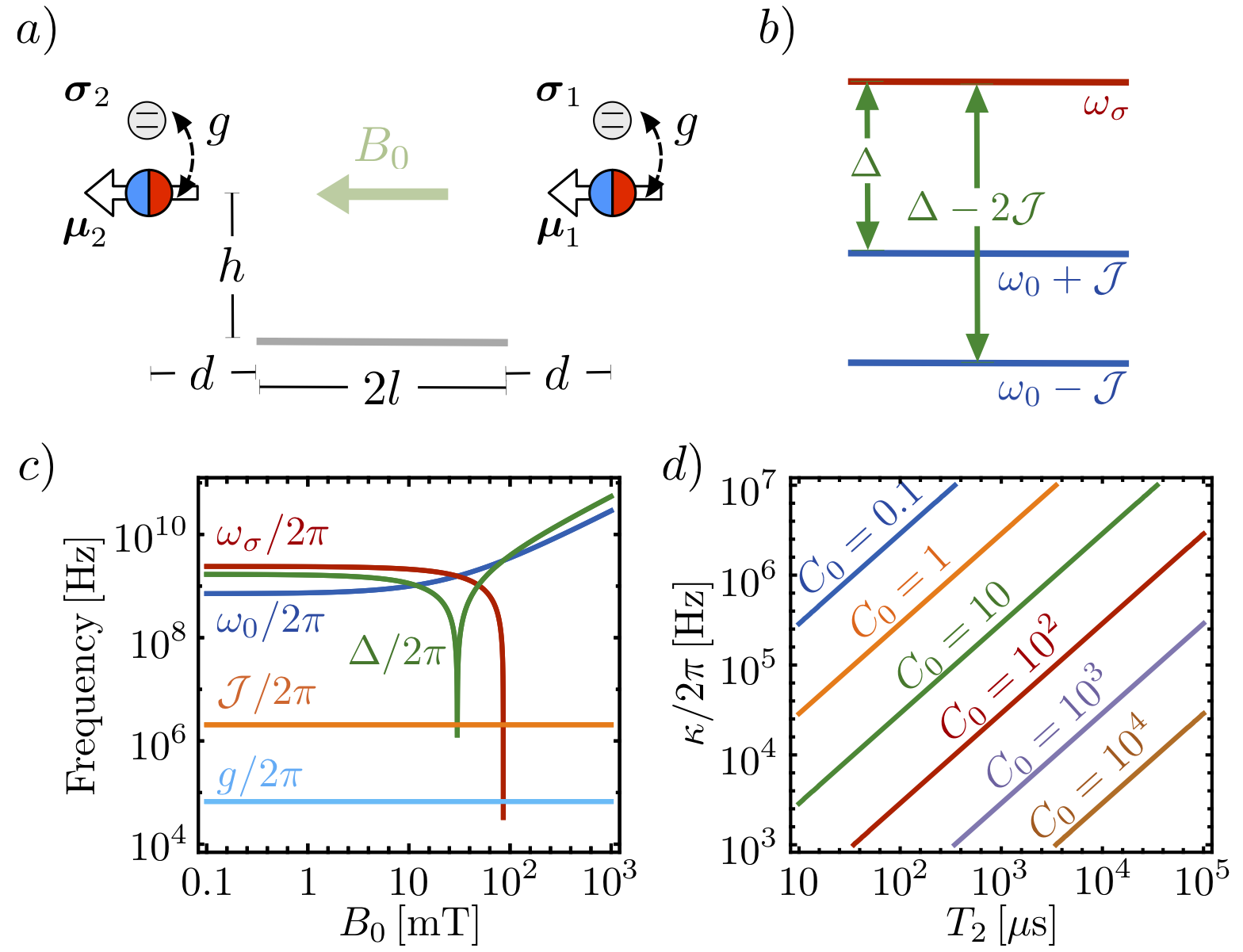}
	\caption{a) Elementary HML configuration (single cell) with two identical spin qubits locally coupled to each magnetic particle. b) Level structure of the system, where $\w_0\pm\mathcal{J}$ are the magnonic normal mode of the elementary cell. c) Relevant frequencies and coupling of the system in a) as a function of the applied field $B_0$. d) Cooperativity $C_0$ as a function of the decoherence rates $\kappa$ and $\pi/T_2$ for a fixed magnet-qubit coupling $g$. Other parameters: $R=350\text{nm}$, $r_q-R=20\text{nm}$, $d-R=100\text{nm}$, $l=5\mum$, $\tau=50\text{nm}$, $h=0$, $\gr=1.76199\times10^{11} \text{rad}\cdot\text{Hz}\cdot\text{T}^{-1}$, $\gq=1.76149\times10^{11} \text{rad}\cdot\text{Hz}\cdot\text{T}^{-1}$, $M_\text{s}=196\times10^3 \text{ T}\cdot\text{m}^{-3}$ (YIG saturation magnetization), and $k_a=2480\text{ J}\cdot\text{m}^3$ (YIG anisotropy energy density). The parameters for YIG are taken from~\cite{Stancil}.}
	\label{Fig:Scaling} 
\end{figure}

Let us consider two identical spin qubits coupled to the elementary configuration described in \figref{Fig:Scaling}.a, and thus separated by a distance $2(d+l)$.
The system is described by the Hamiltonian $\Hop_\text{T} = \sum_{j=1}^2[\hbar \w_0 \fdop_j\fop_j+\mathcal{J}\sum_{i\neq j} \fdop_i\fop_j+\Hop_\text{MQ}^{j}]$. In the dispersive regime, when the qubits are detuned from the magnonic eigenmodes of the system, it is possible to adiabatically eliminate the magnonic degrees of freedom. The qubit dynamics are thus described by the following effective spin-spin interaction Hamiltonian (see~\appref{apdx:SpinSpin_Interaction})
\be\label{eq:HQQ}
\begin{split}
\Hop_\text{QQ} =& \frac{\hbar}{2}\spare{\wsig - \frac{g^2}{2\Delta}-\frac{g^2}{\Delta-2\mathcal{J}}}\pare{\sigz_1+\sigz_2}\\
& - \hbar g_\text{eff}(\sigp_1\sigm_2+\sigm_1\sigp_2),
\end{split}
\ee
where $\Delta\equiv \w_0+\mathcal{J}-\wsig$, and the effective spin-spin coupling strength reads $g_\text{eff}= g^2[1/\Delta -1/(\Delta-2\mathcal{J})]$. The level structure and typical values of frequencies and couplings are shown in \figref{Fig:Scaling}.b-c. $\Hop_\text{QQ}$ can be used to swap excitations between the two qubits at a rate $\pi/\geff$ whenever $\geff \gg \gamma,\keff$, where $\gamma \equiv \pi/T_2^*$ and $\keff = \kp g^2[1/\Delta^2+1/(\Delta-2\mathcal{J})^2]$ is the qubit damping induced by the lossy magnonic bus (see~\appref{apdx:SpinSpin_Interaction})~\footnote{We remark that the effective description in \eqnref{eq:HQQ} breaks down when the qubit is resonant with the magnonic modes, \ie for $\Delta\simeq 0, 2\mathcal{J}$ (see also~\figref{Fig:Scaling}.b). In this cases the dynamics of the system is correctly described by a master equation for the total magnons-qubit system (see~\appref{apdx:SpinSpin_Interaction} for details).}. In this strong coupling regime, the error $\ve$ on the state transfer fidelity for optimized values of the detuning $\Delta$ and magnon-tunneling $\mathcal{J}$ is given by $\ve \approx \sqrt{\alpha_\kp\alpha_{\gamma}/(2C_0)}$ with cooperativity $C_0\equiv g^2 /(\gamma \kp)$ where we numerically estimate $\alpha_{\gamma} \simeq 0.779 $ and $\alpha_\kp \simeq 0.006$ as detailed in~\appref{apdx:SpinSpin_Interaction}. In \figref{Fig:Scaling}.d, values of $C_0$ are shown as a function of magnon damping $\kp$ and qubit dephasing times $T_2$ and fixed values for the remaining parameters. As qubit dephasing times $T_2\approx 0.5\text{s}$ are achievable with dynamical decoupling schemes~\citep{Abobeih2018}, the main limitation is given by the magnon damping rate \citep{Tabuchi2014} for the current experimental state of the art. 

\section{Conclusions}\label{sec:Conclusions}

In conclusion, we have shown that hybrid magnetic lattices allow to implement artificial magnonic crystals with engineered band structures. 
Our approach extends the range of magnetic interactions beyond the limit set by free-space magnetic-dipole interactions and provides an attractive alternative to existing methods, where magnonic crystals are built from arrays of dipolarly coupled nanostripes of magnetic materials~\cite{Qin2018,Topp2010,Gubbiotti2007}.
Furthermore, it presents an alternative platform to study magnetic crystallization and dynamics of low density ensembles of nanomagnets embedded in a non-magnetic matrix. Thus, it is relevant for the field of artificial spin systems~\citep{Nisoli2013,Heyderman2013}.
For those systems it would be interesting to replace the lattice of loops with a superconducting wire network~\citep{Ling1996,Berger2001}, since this would allow to study how the interplay between connectivity and superconductivity affects the dynamics of magnetic particles in the network. 
In addition, spin qubits coupled to the magnets in the network allow to perform local magnetometry and thereby probe the state of the network. The spin network configuration also allows to use magnons as a quantum bus to magnetically couple spin qubits over long distances~\citep{Trifunovic2013}, analogously to what is done with quantum emitters coupled to photonic crystals~\citep{Chang2018}, albeit in a different parameter regime. 

The potential of our proposal depends very much on the linewidth of magnons in a magnetic sphere. 
While the microscopic origin of such damping is still not completely understood, interesting strategies to possibly reduce the damping can be envisioned. 
Smaller magnetic particles might show a lower damping at $T\gtrsim 1K$ due to the discretization of phononic modes in the sample. 
A levitated version of our proposal~\citep{PratCamps2017,Rusconi2017b} might allow to study the impact of the conservation of total angular momentum on the (dissipative) dynamics of the magnetization.
Finally, we remark that the present discussion could be generalized beyond the macrospin approximation to include other magnonic modes inside the magnetic particles which might result in an improvement on the magnon linewidth~\citep{Goryachev2014}.

We thank  J.~I.~Cirac, D.~De Bernardis, J.~J. Garcia-Ripoll,  G.~Kirchmair, K.~Lehnert, C.~Navau, J.~Prat-Camps, P.~Rabl, and A.~Sanchez for useful discussions.
We thank M.~Juan for support with graphical illustrations.
CCR and ORI aknowledge support from the European Research Council (ERC-2013-StG 335489 QSuperMag) and the Austrian Federal Ministry of Science, Research, and Economy (BMWFW).
MJAS thanks the Humboldt foundation for financial support.
JG acknowledges support from the European Union (SEQOO, H2020-MSCA-IF-2014, no. 655369).

CCR and MJAS contributed equally to this work.


\appendix

\section{Description of a superconducting loop}\label{apdx:ModelLoop}

In the following, we describe a superconducting loop as a multimode microwave resonator, and we derive under which conditions it can be approximated as a single mode LC-oscillator.

Superconducting rings on top of a dielectric substrate have been shown to behave as microwave multimode resonators~\citep{Chang2004,Hopkins2008} characterized by a large quality factor $Q\approx 10^6$ at $\text{GHz}$ frequencies~\citep{Minev2013,Minev2016,Paik2011}.
The spectrum of the resonator is double degenerate, each frequency corresponding to both a clockwise and counter-clockwise traveling wave.
Within a transmission line model the mode frequencies can be approximated by $\w_n/2\pi \equiv n/(2\pi l \sqrt{L_lC_l})$ for $n\in \mathbb{N}$, where $L_l$ ($C_l$) is the inductance (capacitance) per unit length of the loop and $l$ is the loop radius. 

Adjusting the external magnetic field $B_0$ such as to tune the Larmor precession frequency of the magnetic particle's macrospin close to the fundamental resonance of the ring resonator, it is possible to neglect the coupling between $\FFop$ and the higher resonant modes. Moreover the degeneracy of the fundamental mode can be broken by introducing small asymmetries or imperfections as done for instance in~\citep{Minev2013,Minev2016}. The ring thus behaves as a single mode $LC$-resonator of frequency $\w_c\equiv 1/\sqrt{LC}$, where $L$ ($C$) is the total inductance (capacitance) of the ring. $C$ is the capacitance between the loop and the ground plate at the opposite end of the dielectric substrate, and can be arbitrarily reduced by careful design. $L$ amounts to the geometrical self-inductance of the loop, which depends on the particular shape of the loop and on the thickness $\tau$ of the wires as detailed in~\citep{Grover2004}. For the case of a circular loop of radius $l$ and wire of circular section, the self-inductance reads~\citep{Grover2004}
\be\label{eq:L}
	L = \mu_0 l \spare{\ln\pare{\frac{8l}{\tau}}-\frac{7}{4} + O\pare{\frac{\tau^2}{l^2}}}. 
\ee
Here we are assuming for simplicity the electric permittivity (magnetic permeability) of the substrate supporting the loop, \figref{Fig:d_crit}.a,  to be $\ve_\text{r}\approx 1$ ($\mu_\text{r} \approx 1$).

\section{Derivation of the system Hamiltonian}\label{apdx:Ham_Derivation}

In the following, we derive the quantum mechanical Hamiltonian \eqnref{eq:Hamiltonian} describing the dynamics of the coupled system composed by the circuit and the magnetic dipole moments.

Within the single mode approximation, a superconductive $LC$-ring resonator can be modeled as an $LC$-circuit (see~\appref{apdx:ModelLoop}). The equations of motion for the $LC$-circuit can be derived from Kirchhoff's current and voltage laws, together with the constitutive relations which relate current and voltage at each element of the circuit. Defining $V_C\equiv \pa{t}\Phi_C$ ($V_L\equiv\pa{t}\Phi_L$) the flux at the capacitor (inductor) of the circuit, we write the constitutive relations for the capacitor as $\Ddot{\Phi}_C = I_C/C$ and for the inductor as
\be\label{eq:L_ConstRel}
	\Phi_L = L I_L + \sum_{j=1}^N \Phi_j(\mmu_j).
\ee
Here, $C$ ($L$) are the circuit capacitance (inductance), and $\Phi_j(\mmu_j)$ is the flux induced in the ring by the $j$-th magnetic dipole.
The equation of motion for the circuit can be derived from Kirchhoff's law as
\be\label{eq:LC_EoM}
	C\Ddot{\Phi} + \frac{\Phi}{L} = \sum_{j=1}^N\frac{\Phi_j(\mmu_j)}{L},
\ee
where $\Phi \equiv \Phi_L = -\Phi_C$~\footnote{Without loss of generality, we assumed no flux trapped in the loop at the initial time.}.

The coherent dynamics of the magnetic moment $\boldsymbol{\mu}_j\equiv \mu (\cos\vphi_j\sin\theta_j,\sin\vphi_j\sin\theta_j,\cos\theta_j)$, for $\theta_j\in[0,\pi]$ and $\vphi_j\in[0,2\pi]$,  is described by the Landau-Lifshitz equation $\pa{t}\mmu_j = -\gr \mmu_j\times \BB(\rr_j)$, where $\BB(\rr_j)$ is the total magnetic field acting on the $j$-th magnetic moment. 
In term of the polar $\vphi_j$ and azimuthal $\theta_j$ angles the Landau-Lifshitz equations read~\citep{Miltat}
\be\label{eq:NM_EoM}
\begin{split}
	\Dot{\vphi}_j =& -\frac{\gr}{\mu\sin\theta_j}\pa{\vphi_j}U,\\
	\Dot{\theta}_j =& \frac{\gr}{\mu\sin\theta_j}\pa{\theta_j}U,
\end{split}
\ee
where $U\equiv \sum_{j=1}^N V^j_a(\mmu_j)+U_0+U_\text{d} + U_\text{ind}$ is the magnetic interaction energy of the dipoles. $V^j_a(\mmu_j)$ represents the magnetic anisotropy energy of the $j$-th magnetic particle.
$U_0 = -\sum_{j=1}^N \mmu_j\cdot \BB_0$ represents the interaction energy of the dipoles with the external bias field. $U_\text{d}$ represents the free-space dipole-dipole interaction energy between the magnetic moments
\be
	U_\text{d}(\{\mmu_j\}) =-\frac{1}{2}\sum_{j=1}^N\sum_{i\neq j=1}^N \mmu_j\cdot \BB_i^\text{dip}(\rr_j).
\ee
The dipolar field created by the dipole moment $\mmu_i$ at position $\rr$ reads
\be\label{eq:BdipNM}
	\BB_i^\text{dip}(\rr)=  \frac{\mu_0}{4\pi}\spare{\frac{3 \Delta\rr_i (\mmu_i\cdot\Delta\rr_i)}{\vert\Delta\rr_i\vert^5}-\frac{\mmu_i}{\vert\Delta\rr_i\vert^3}},
\ee
with $\Delta \rr_i\equiv \rr-\rr_i$.
$U_\text{ind} = I_L\sum_j \Phi_j(\mmu_j)$, where $I_L$ is given by \eqnref{eq:L_ConstRel}, represents the interaction between the magnetic dipoles and the field produced by the current flowing in the ring~\citep{Jackson}.

The equations of motion \eqnref{eq:LC_EoM} and \eqnref{eq:NM_EoM} can be derived from the Lagrangian
\be\label{eq:Lagrangian}
\begin{split}
	\mathcal{L}=&  \frac{C}{2}\Dot{\Phi}^2\! +\frac{\mu}{\gr}\sum_{j=1}^N\Dot{\vphi}_j\cos\theta_j\!-\!\frac{1}{2L}\bigg[\Phi\!-\!\sum_{j=1}^N\Phi_j\big(\vphi_j,\theta_j\big)\bigg]^2\\
	&-\frac{1}{2L}\!\bigg[\sum_{j=1}^N\Phi_j\big(\vphi_j,\theta_j\big)\!\bigg]^2\!-\! \sum_{j=1}^N \mu B_0 \cos\theta_j - U_\text{dip}\\
	&-\sum_j V^j_a(\vphi_j,\theta_j).
\end{split}
\ee
From \eqnref{eq:Lagrangian}, the classical Hamiltonian of the system is obtained introducing the generalized momenta $Q\equiv C\Dot{\Phi}$ and $p_j \equiv \mu \cos\theta_j/\gr$ conjugated to $\Phi$ and $\vphi_j$ respectively.
Following the usual canonical quantization procedure one can then derive the quantum mechanical Hamiltonian of the system given in \eqnref{eq:Hamiltonian}.
The charge $\Qop$ and flux $\Phop$ operators of the circuit appearing in the system Hamiltonian satisfy canonical commutation relations $[\Phop,\Qop]=\im \hbar$. The components of the magnetic moment $\mmop_j \equiv \mu(\sin\hat{\theta}_j\cos\hat{\vphi}_j,\sin\hat{\theta}_j\sin\hat{\vphi}_j,\cos\hat{\theta}_j)^T$, commute as $[\muop_j^\nu,\muop_i^\eta]= \im \mu \delta_{ij}\levi{\nu\eta\xi}\muop_j^\xi$, for $\nu,\eta,\xi=x,y,z$, according to the canonical quantization of the classical Poisson bracket
\be
	\cpare{f,g} = -\sum_{j=1}^N \frac{1}{\mu\sin\theta_j}\pare{\frac{\pa{}f}{\pa{}\vphi_j}\frac{\pa{} g}{\pa{}\theta_j}-\frac{\pa{}f}{\pa{}\theta_j}\frac{\pa{} g}{\pa{}\vphi_j}},
\ee
for any $f,g$ functions of $\theta_j,\vphi_j$.

\section{Magnetic flux through a coil}\label{apdx:FluxCoil}

In the following, we derive the expression for the flux induced by a magnetic dipole moment in a neighbouring superconducting loop.

We consider the inductive coupling between a magnet with magnetic moment $\hat{\boldsymbol{\mu}}=\hbar \gr \FFop$ and a coil of arbitrary shape. We assume the magnet to be placed at a distance $h$ above the coil and at a horizontal distance $d$ from the coil's closest wire (see~\figref{Fig:Illustration}.b in the main text).
The flux induced in the coil by the magnet reads
\be\label{eq:Def_Flux}
	\Phi(\FFop) = \oint \text{d}\bold{l}\cdot \AB(\rr, \FFop)
\ee
where $\AB(\rr, \FFop)$ is the magnetic vector potential generated by the magnet and the integral is taken on the contour defined by the coil. 
\eqnref{eq:Def_Flux}  can be written as $\Phi(\FFop)= \hbar \gr\mu_0\sum_\nu I_\nu \Fop_\nu/(4\pi d)$ for $\nu=x,y,z$ where
\be\label{eq:I_factor}
	I_\nu \equiv d \oint \frac{(\Delta\rr \times \text{d}\bold{l})_\nu}{\vert\Delta\rr\vert^3},
\ee
is a dimensionless factor which depends only on the shape of the coil and on the mutual position of the magnet and the coil. Here $\Delta\rr$ is the distance between the magnet and a point in the coil.
For instance, for a circular coil of radius $l$ centered at $(0,0,l+d)$ and for a nanomagnet at a position $(h,0,0)$, the factors $I_\nu$ in \eqnref{eq:I_factor} read
\be\label{eq:Icirc}
\begin{split}
	I_z =& \int_{-l/d}^{l/d} \text{d}\,\lambda\, F\pare{\lambda,\frac{l}{d},\frac{h}{d}},\\
	I_x=& \int_{-l/d}^{l/d} \text{d}\lambda \,G\pare{\lambda,\frac{l}{d},\frac{h}{d}},
\end{split}
\ee
and $I_y=0$, where
\be
\begin{split}
	F(\lambda,x,y)\equiv &\frac{y\lambda/\sqrt{x^2-\lambda^2}}{\spare{y^2 + x^2+\pare{x+1}+2\lambda \pare{x+1}}^{3/2}},\\
	G(\lambda,x,y)\equiv &\frac{\sqrt{x^2-\lambda^2}+\frac{\lambda}{\sqrt{x^2-\lambda^2}}\pare{x+1+\lambda}}{\spare{y^2 + x^2 + \pare{x+1}+2\lambda \pare{x+1}}^{3/2}}.
\end{split}
\ee

\begin{figure}
	\includegraphics[width=0.8\columnwidth]{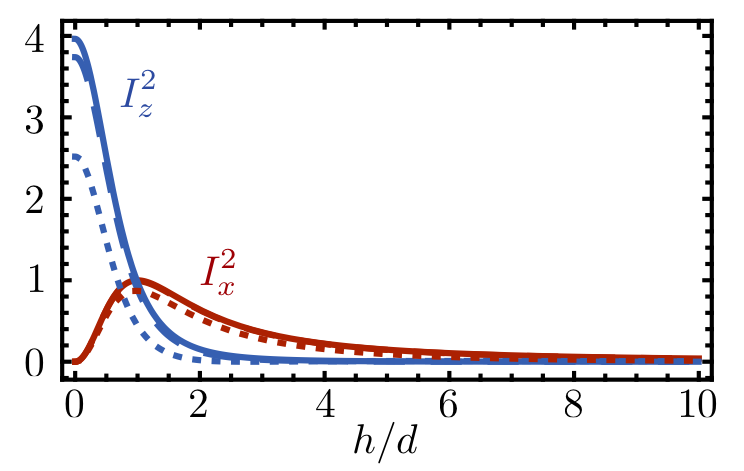}
	\caption{Plot of the parameters $I_z$ and $I_x$ in function of $h/d$ for $l/d=10$ (dotted line), $10^2$ (dashed line), $10^3$ (solid line). For $I_x$ the dashed and solid line are almost coincident.}
	\label{Fig:I_factor} 
\end{figure}

As shown in \figref{Fig:I_factor}, the integrals in \eqnref{eq:Icirc} have an optimal value around unity in function of $h/d$ in the limit of a large loop radius, $l/d\gg 1$.
%

\section{Magnon dynamics in a HML: Hamiltonian derivation}\label{apdx:Hamiltonian_SLNM}

In the following, starting from \eqnref{eq:Hamiltonian}, we derive \eqnref{eq:QuadHam}, which describes the propagation of magnons in a HML. We consider the simple case of two magnets as depicted in \figref{Fig:Illustration}.b (generalization to the case of several magnets is straightforward).
By substituting the definition \eqnref{eq:Phi_j} into \eqnref{eq:Hamiltonian}, one obtains
\be\label{intermediate_Ham}
\begin{split}
	\Hop =& \hbar \w_c \adop\aop-\frac{\Phi_c}{L}(\adop+\aop) \delta\Phop +\hbar \wL\sum_{j=1}^2\Fop_j^z+\frac{(\delta\Phop)^2}{L}\\
	& + \frac{\Phi_\text{bias}}{L}\delta\Phop + \hat{U}_\text{d}(\{\mmop_j\})+\sum_{j=1}^2\Vop_a^j(\mmop_j),
\end{split}
\ee
where $\w_c\equiv 1/\sqrt{LC}$, $\Phi_\text{bias} \equiv \sum_j \Phi_\text{bias}^j$, and $\delta\Phop\equiv \sum_{j}\delta\Phi_j(\Delta\mmop_j) $. Here, we expressed the circuit operators in terms of creation and annihilation operators $(\Phop-\Phi_\text{bias})\equiv\Phi_c(\adop+\aop)$ and $\Qop \equiv \im(\adop-\aop)/(2\Phi_c)$, where $\Phi_c\equiv \sqrt{\hbar/(2C\w_c)}$.

We consider the applied field $\BB_0$ to be sufficiently large as to initially polarize the macrospin at the two nodes along $-\uez$, such that $\avg{\FFop_j}_0 = -F \uez$. The fluctuations of $\FFop_j$ around the equilibrium state can be described by a bosonic mode $\fop_j,\fdop_j$ (magnon) according to the Holstein-Primakoff approximation $\Fop_j^z = -F + \fdop_j\fop_j$, and $\Fop_j^+ \simeq \sqrt{2F}\fdop_j$.
In the limit of small fluctuations $\avg{\fdop_j\fop_j}\ll 2F$, $\Hop$ can be approximated by a quadratic Hamiltonian in the bosonic operators $\adop,\aop, \fdop_j$ and $\fop_j$ as
\be\label{eq:Hm_Quad}
\begin{split}
	\Hop \simeq & \hbar \w_c\adop\aop-\hbar \pare{\adop+\aop}\sum_{j=1}^2\pare{\chi_j\fop_j + \hc}\\
	&+ \hbar \sum_{j=1}^2 \Big\{\w_j\fdop_j\fop_j + 2\sum_{i}\pare{\Lambda_{ij} \fdop_i\fdop_j +\hc}\\
	&+ \sum_{i,j=1}^2\big[(2-\delta_{ij})\mathcal{J}_{ij}+(1-\delta_{ij})\mathcal{J}_{ij}^\text{d}\big]\fdop_i\fop_j\Big\}\\
	&+\hbar \sum_{j=1}^2(\eta_j \fop_j+\eta^*_j\fdop_j).
\end{split}
\ee
We have defined
	\bea
	\w_j&\equiv& \gr B_0 + 2\frac{\gr k_a}{M_\text{s}} + \mathcal{J}_{jj}\nonumber\\ 
	& &+ \sum_{i=1}^2\frac{\hbar \gr^2\mu_0}{4\pi r_{ij}^3}(3\cos^2\!\theta_{ij}- 1\!)F,\qquad\label{eq:w_0}\\
	\Lambda_{ij} &\equiv & -3(1-\delta_{ij})\frac{\hbar \gr^2\mu_0}{16\pi r_{ij}^3}	F\sin^2\theta_{ij}e^{\im2\vphi_{ij}}\nonumber\\
	& & +\frac{\Phi_\text{e}^2F}{2\hbar L}I_jI_i,\label{eq:Lamda}\\
	\mathcal{J}_{ij} &\equiv &\pare{\frac{\hbar \gr \mu_0}{4\pi}}^2\frac{I_{ij} F}{2\hbar d_i d_j L},\label{eq:Jij}\\
	\mathcal{J}_{ij}^\text{d} &\equiv &  -\frac{\hbar \gr^2 \mu_0}{8\pi r_{ij}^3}(3\sin^2\theta_{ij}-2)F,\label{eq:JijDip}\\
	\chi_j &\equiv& \frac{\Phi_\text{e}\Phi_c}{2\hbar L} I_j\sqrt{2F}\label{eq:Chi},\\
	\eta_j &\equiv& \sum_i\frac{3\hbar \mu_0\gr^2}{8\pi r_{ij}^3}F\sqrt{2F}e^{\im2\vphi_{ij}}\cos\theta_{ij}\sin\theta_{ij}\nonumber\\
	& & +\frac{\Phi_\text{e}\Phi_\text{bias}}{2\hbar L}\sqrt{2F}I_j.\label{eq:eta}
\eea
Here, $\Phi_e$ is independent of $j$ as we assumed the particles to be at the same distance $d$ from the loop's wire (see~\secref{SSec:Hamiltonian}), $k_a$ is the magnetic anisotropy energy density and $M_\text{s}$ is the saturation magnetization of the magnetic particle. We additionally assumed the easy magnetization axis of the magnetic anisotropy potential of the material to be aligned along the direction of the applied magnetic field. In this case, the anisotropy energy contributes only as a shift to the magnon oscillation frequency within the quadratic approximation.

The linear term in \eqnref{eq:Hm_Quad} shifts the equilibrium orientation of the magnetic moments and the equilibrium value of the flux in the loop. It can be formally eliminated from \eqnref{eq:Hm_Quad} displacing the bosonic operators $\adop$, $\aop$, $\fdop_j$, and $\fop_j$ to represent the fluctuations around the new equilibrium values. 
The linear term in \eqnref{eq:Hm_Quad} is identically zero when the magnetic particles are placed in the plane of the LC-resonator ($h=0$), as in \figref{Fig:d_crit}.a, and the distance between the magnets is such that the free-space dipole-dipole interaction is negligible~\footnote{For the elementary two-magnet configuration in \figref{Fig:Illustration}.b  $\theta_{12}=0, \vphi_{ij}=\pi/2$ thus the first term in \eqnref{eq:eta} cancels.}. 
We thus neglect hereafter the last term in \eqnref{eq:Hm_Quad} assuming the shift in the relevant couplings and frequencies to be negligible ($h\sim 0$). We remark that all the quantitative predictions made in the main text are calculated for $h=0$ and negligible dipole-dipole interaction.

Due to the large detuning between $\w_c$ and the frequencies defined in Eq.~(\ref{eq:w_0}-\ref{eq:Chi}), we adiabatically eliminate the $LC$-resonator degrees of freedom, which are assumed to be in the vacuum state.
Within the rotating-wave approximation and taking into account the circuit-induced shifts of the frequencies and couplings, one obtains the effective Hamiltonian \eqnref{eq:QuadHam} that describes the magnon dynamics.

\section{Magnetic field intensity at the wires of the loop}\label{apdx:critical_distance}

Here, we calculate the field produced by the magnetic particle at the wire of the loop. From the requirement that this field should not exceed the critical field to keep the loop in the Meissner state, we derive the minimal distance from the wire at which a magnetic particle can be placed.

\begin{figure}[H]
	\includegraphics[width=0.9\columnwidth]{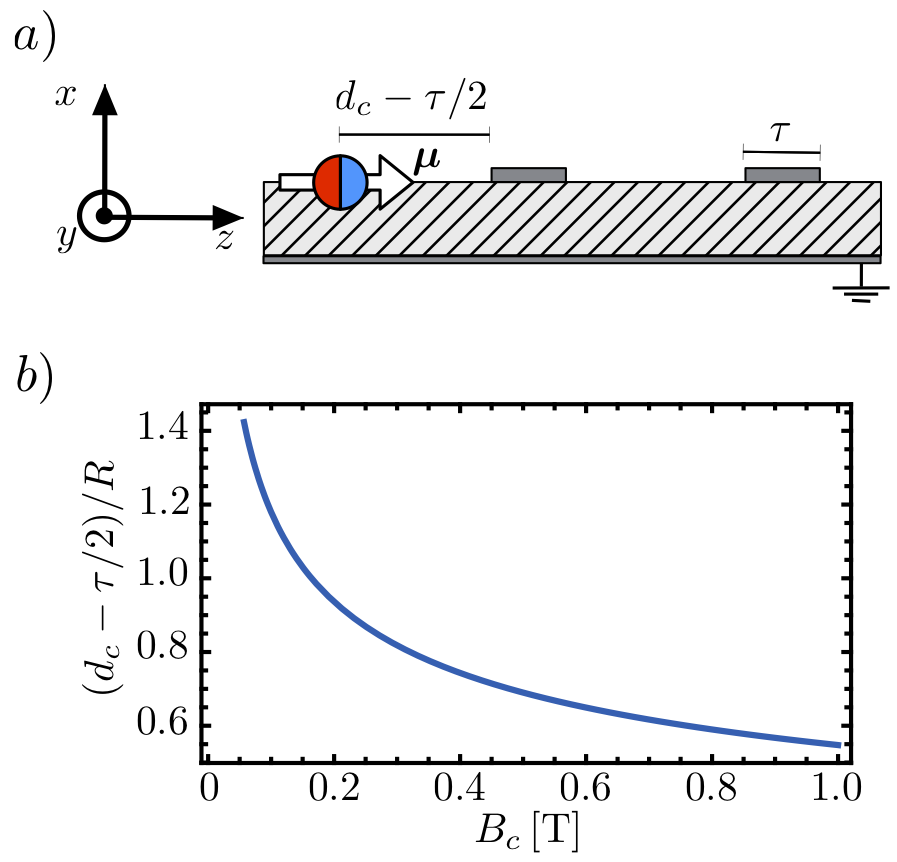}
	\caption{a) Lateral view (not to scale) of a magnetic particle positioned on the substrate (hatched grey region) sustaining the superconducting-ring ($h=0$) at a distance $d-\tau/2$ from the closest point of the superconducting-ring wire (dark grey region) of thickness $\tau$. b) Critical distance $d_c-\tau/2$ in unit of $R$ at which the field produced by the magnetic moment at the closest wire point equals $B_c$. Other physical parameters are taken from the caption of \figref{Fig:Scaling}.}
	\label{Fig:d_crit} 
\end{figure}

The magnetic field produced by the particle at any point in the wire must be smaller than the critical field (first crytical field) $B_c$ of the type I (type II) superconductor that makes up the loop. 
Consider the situation illustrated in \figref{Fig:d_crit}.a. The distance at which the center of the magnetic particle should be placed such that the $\uez$-component of the magnetic field at the closest point of the loop equals $B_c$ reads
\be\label{eq:d_crit}
	d_c = \frac{\tau}{2}+\pare{\frac{2\mu_0 M_\text{s}}{3B_c}}^{1/3}R,
\ee
where $M_\text{s}$ is the saturation magnetization of the magnetic particle, $R$ the particle radius, and $\tau$ the wire thickness.
In \figref{Fig:d_crit}.b, $(d_c-\tau/2)/R$ is plotted as a function of the field $B_c$ at the wire position.
For the values used in \figref{Fig:Scaling}, the field produced by the magnetic particle at the position of the loop wire is $\approx 110\text{mT}$, which is below the first critical field of many type II superconductor such as Nb~\citep{Roberts1976}.
%

\section{Bone-shape configuration}\label{apdx:Bone}

\begin{figure}
	\includegraphics[width=0.9\columnwidth]{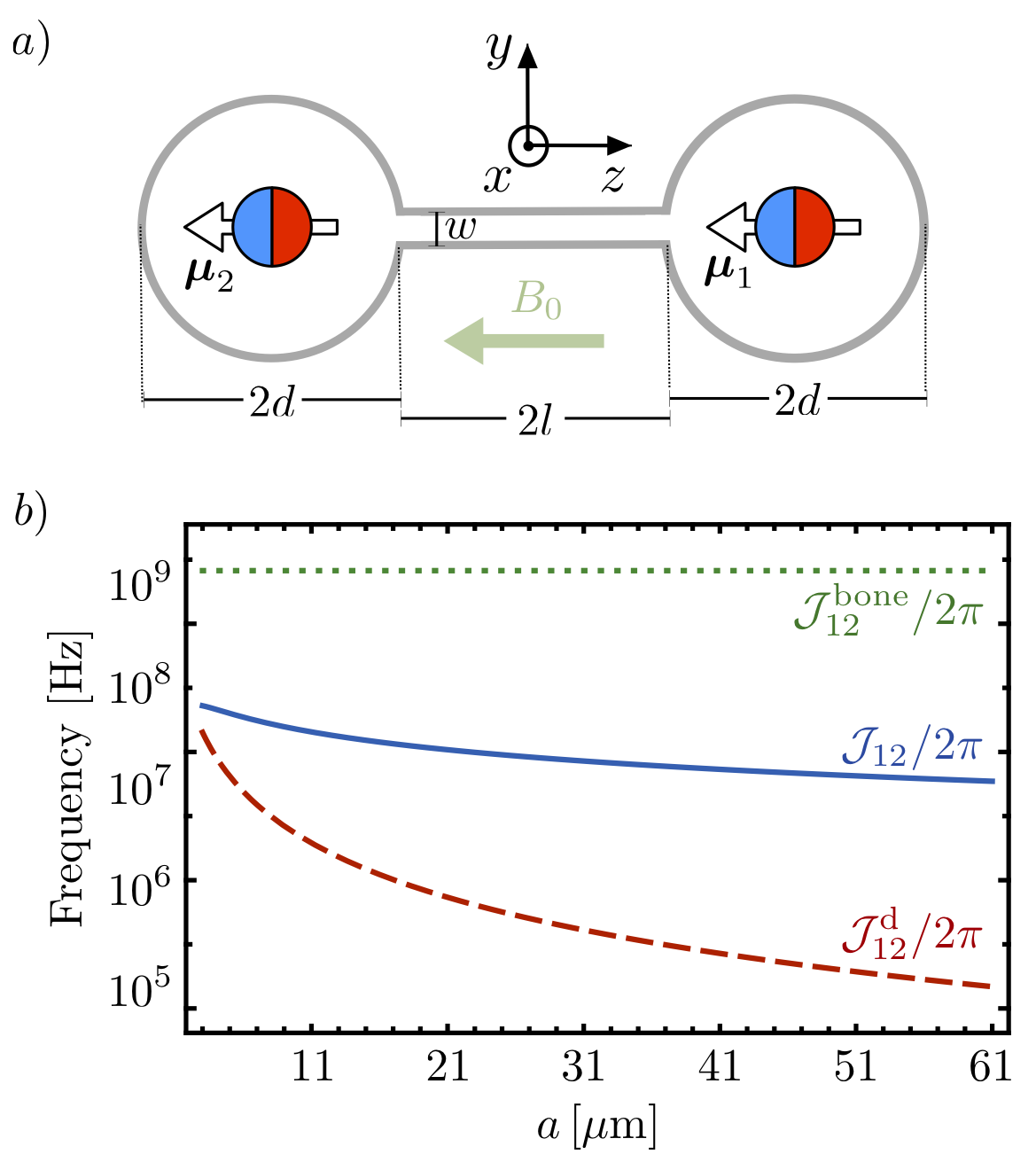}
	\caption{a) Bone shape coil for enhancing the magnon tunneling rate between distant magnets. In such a configuration the total distance between the magnets is $a=2d+2l$. b) Magnon tunneling rate $\mathcal{J}_{12}$ (solid blue line),  $\mathcal{J}_{12}^\text{d}$ (dashed red line), and $\mathcal{J}_{12}^\text{bone}$ (dotted green line) as function of the magnets separation $a$. We assumed $R=1\,\mu\text{m}$, $d=500\,\text{nm}$, and $\tau = 50\,\text{nm}$. The tunneling rate $\mathcal{J}_{12}$ ($\mathcal{J}^\text{d}_{12}$) corresponds to the loop-mediated (free-space magnetic dipole) tunneling rate in the configuration shown in \figref{Fig:Illustration}.b.}
	\label{Fig:Bone} 
\end{figure}

In the following, we analyze a different coil geometry in which the magnets are inscribed inside the perimeter defined by the coil's wire. We show that while such a configuration is not suitable to build hybrid magnetic lattices, it can achieve a larger magnon tunneling rate than the configuration in \figref{Fig:Illustration}.b.
Let us consider the situation illustrated in \figref{Fig:Bone}.a, where two magnets are coupled through a bone-shaped loop.
Here, $d$ is the radius of the circular end-rings, $w$ the separation of the middle parallels wires, and $2l$ their length.
For $w\ll d^2/l$, the middle region connecting the two circular ends of the loop has a negligible contribution to the self inductance $L$ of the loop. Moreover for $R<d$, the magnetic flux produced by the magnetic particle is obtained as the flux generated by a magnetic moment $\boldsymbol{\mu}$ placed at the origin of a circular coil and of intensity $\mu = M_\text{s}4\pi R^3/3$. For $l\gg d$, the flux produced by a magnet in the loop at the opposite end of the coil can be neglected.
The fluctuating magnetic moment $\mmop_j$ produces a fluctuating flux $\Phop_j = \hbar \gr \mu_0\Delta \Fop_x/(4l)$, where $\Delta \Fop_y$ and $\Delta\Fop_z$ contribute only at higher order.
In this configuration, the direct inductive magnetic coupling contribution to the  magnon-tunneling rate thus reads
\be\label{eq:J_bone}
	\mathcal{J}_{12}^\text{bone} \equiv \frac{ \gr^2\mu_0^2 \hbar F}{8d^2L}.
\ee
Here, $L$ represents the inductance of a circular coil of radius $d$ [cf. \eqnref{eq:L}].

In \figref{Fig:Bone}.b, $\mathcal{J}_{12}^\text{bone}$ is plotted as function of the magnet separation $a=2(l+d)$, keeping $d$ fixed. Realistically, at larger values of $l$ the contribution of the middle region to the total inductance will affect the scaling of $\mathcal{J}_{12}^\text{bone}$. However, for a sufficiently small separation $w$ between the two parallel wires the tunneling rate is expected to vary only slightly with an increase of $l$. The bone-shape configuration (\figref{Fig:Bone}.a) thus allows to enhance the magnon tunneling rate for a given separation $a$ as compared to the simpler configuration in \figref{Fig:Illustration}.b.
%

\section{Hamiltonian for the 2D HML}\label{apdx:2D_SLNM}

In the following, we derive the Hamiltonian \eqnref{eq:Ham2D} which describes the two-dimensional HML illustrated in \figref{Fig:Schematic}.d, whose band structure is shown in \figref{Fig:Schematic}.e. 

Let us consider the elementary cell of such a configuration shown in \figref{Fig:Schematic}.c.
For large loop size $l$, $\mathcal{J}_{ij}^d\ll\mathcal{J}_{ij}$ (see~\figref{Fig:Bone}.b), and thus the magnon tunneling rate is given by $\mathcal{J}_{ij} = \mathcal{J}\,e^{\im \phi_{ij}}$, with 
\be
	\mathcal{J} = \frac{\Phi_\text{e}^2 F}{2\hbar L}(I_x^2+I_y^2),
\ee
where we defined $I_x \equiv I_1^x = I_2^x=I_3^x=I_4^x$ and $I_y \equiv I_1^y = I_3^y = -I_2^y = -I_4^y$, and $\phi_{ij}$ is a function of $I_x$, $I_y$.
The magnon dynamics in the elementary cell in~\figref{Fig:Schematic}.c is thus described by 
\be
	\Hop_\text{M}^\text{cell} = \hbar \w_0 \sum_{j=1}^4 \fdop_j\fop_j + \hbar \mathcal{J} \sum_{i\neq j=1}^4 \fdop_i\fop_j,
\ee
where we redefined some of the magnonic operators to absorb the phase factor appearing in the tunneling rate $\mathcal{J}_{ij}$.

The extended 2D HML shown in \figref{Fig:Schematic}.d is built by repetition of this elementary cell. As discussed in the main text, the magnon dynamics of such a 2D HML can be described by a two interacting sublattices model, labelled by $A$ and $D$ according to the Hamiltonian
\be\label{eq:Ham2D_SM}
\begin{split}
    \Hop_\text{M}^{2\text{D}} =&\hbar \w_0 \sum_\bj \pare{ \fop^{\text{D}\dag}_\bj\fop^\text{D}_\bj + \fop^{\text{A}\dag}_\bj \fop^\text{A}_\bj }\\
    &+ \hbar\mathcal{J}\bigg[\sum_{\bj,\bbeta} \fop^{\text{D}\dag}_\bj \fop^\text{A}_{\bj+\bbeta}+\sum_{\bj,\balp} \fop^{\text{A}\dag}_\bj\fop^\text{A}_{\bj+\balp}\\
    &+\sum_{\bj,\bdlt} \fop^{\text{D}\dag}_\bj \fop^\text{D}_{\bj+\bdlt}+\hc \bigg].
\end{split}
\ee
Here, the operators $\fop^\text{A}_\bj,\fop^{\text{A}\dag}_\bj$ ($\fop^\text{D}_\bj,\fop^{\text{D}\dag}_\bj$) respectively create and annihilate a magnon in the sublattice $A$ ($D$) within the cell at position $\bj=(j_y ,j_z)$ and the vectors $\balp$ and $\bbeta$ ($\bdlt$ and $\bbeta$), for $\bbeta \in \{(\pm 1/2,\mp 1),(\pm 1/2,0)\}$, $\balp \in \{(\mp1,\pm 1)\}$, and $\bdlt \in \{(\pm1,0)\}$,  connect the nearest neighbors of a point in the sublattice $A$ ($D$) in the basis of the Bravais vectors $\bold{v}_1$ and $\bold{v}_2$.

In terms of the operators 
\bea
	\fop^\text{D}_\kk &=& \frac{1}{N}\sum_\bj e^{-\im \kk\cdot \bj} \fop^\text{D}_\bj,\\
	\fop^{\text{D}\dag}_\kk &=& \frac{1}{N}\sum_\bj e^{-\im \kk\cdot \bj} \fop^{\text{D}\dag}_\bj,\\
	\fop^\text{A}_\kk &=& \frac{1}{N}\sum_\bj e^{-\im \kk\cdot \bj} \fop^\text{A}_\bj,\\
	\fop^{\text{A}\dag}_\kk &=& \frac{1}{N}\sum_\bj e^{-\im \kk\cdot \bj} \fop^{\text{A}\dag}_\bj,
\eea
which create/annihilate a magnon of momentum $\kk=(k_y,k_z)$ in the sublattice $D$ or $A$, the Hamiltonian in \eqnref{eq:Ham2D_SM} can be written as 
$\Hop_\text{M}^{2\text{D}} = 2\mathcal{J} \Psdop M \Psop$ where $\Psop \equiv (\fop^\text{D}_\kk,\fop^\text{A}_\kk)^T$ and
\be
	M\! \equiv\! \begin{pmatrix}
		2\cos[(k_y+k_z)a] &\! \cos(k_y a)+\cos(k_z a)\\
		\cos(k_x a) + \cos(k_y a)\! & 2 \cos[(k_x-k_y)a]
	\end{pmatrix}\!.
\ee
The eigenvalues of $\Hop_\text{M}^{2\text{D}}$ read
\be\label{eq:Bands}
\w_\pm(\kk)= \w_0 + 2\mathcal{J}[4\cos(k_x a)\cos(k_y a)\pm\sqrt{\Lambda}],
\ee
where $\Lambda\equiv 4+4\cos(k_x a)\cos(k_y a)-\cos(2k_y a)-\cos(k_x a)+2\cos(2k_x a)\cos(2k_y a)$, with $a=\sqrt{2}(l+d)$ being the lattice constant. \eqnref{eq:Bands} correspond to the magnon bands illustrated in \figref{Fig:Schematic}.e.

\section{Spin qubits coupled to a HML}\label{apdx:NV-Magnon}

In the following, we derive derive the general Hamiltonian \eqnref{eq:JCH_kspace} describing NV-center qubits coupled to the magnetic particles in a HML. In particular, we obtain first \eqnref{eq:H_MQ} for an NV-center qubit coupled to a magnetic particle by magnetic dipole-dipole interaction, and we later generalize this result to the case of several NV-center qubits coupled to a HML.

\begin{figure}
	\includegraphics[width=\columnwidth]{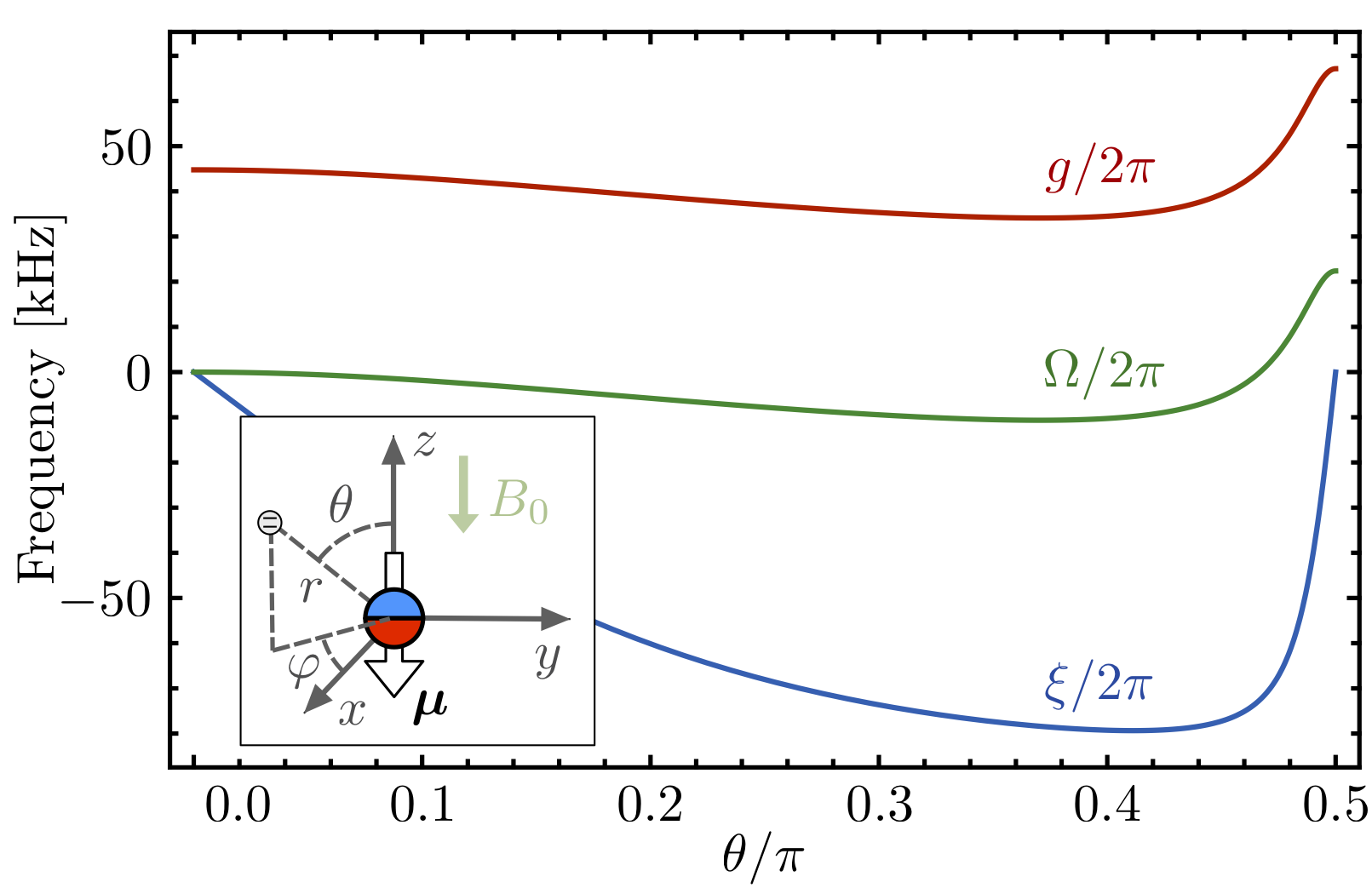}
	\caption{Plot of the couplings $\xi$, $g$, and $\W$ vs the position angle $\theta$. The remaining parameter have the following values: $R=350\text{nm}$, $r_0=R+20\text{nm}$, $B_0=70\text{mT}$, and $\gr$, $\gq$ as in the caption of \figref{Fig:Scaling}. Inset: general configuration of the nanomagnet-qubit system.}
	\label{Fig:Coupling_Theta} 
\end{figure}

Let us first consider a single NV-center located at $\rr_q\equiv r_q(\sin\theta\cos\varphi,\sin\theta\sin\varphi,\cos\theta)$ around a magnetic particle (see inset~\figref{Fig:Coupling_Theta}). The magnetic dipole interaction Hamiltonian between the NV-center and the magnetic particle reads,
\be\label{eq:Ham_NV-Magnet}
	\Hop_\text{M-NV} = \hbar\Delta_\text{NV}\Sop_z^2 + \gq B_0 \Sop_z -\hbar \gq \SSop \cdot \hat{\BB}^\text{dip}(\rr_q),
\ee
where $\SSop$ is spin-1 operator of the NV-center, $\Delta_\text{NV}$ its zero field splitting, $\gq$ its gyromagnetic ratio, and $\BB^\text{dip}(\rr_q)$ is the dipole field produced by the magnet at the NV position [see~\eqnref{eq:BdipNM}]. In the following, as the derivation is the same at each node, we drop the site-index $j$.

Expressing the NV spin operators in terms of the eigenstates of $\Sop_z$, namely $\Sop_z = \ketbra{1}{1}-\ketbra{-1}{-1}$ and $\Sop_+ = (\Sop_-)^\dag = \sqrt{2}(\ketbra{0}{-1}+\ketbra{1}{0})$, the Hamiltonian \eqnref{eq:Ham_NV-Magnet} can be rewritten as $\Hop_\text{M-NV}=\Hop_1+\Hop_{-1}$, where $\Hop_k$ acts only on the states $\ket{0},\ket{k}$ ($k=\pm1$) of the NV center. 
In the following, we assume the frequency of the magnon to be close to the NV center transition frequency between $\ket{0}$ and $\ket{-1}$. This can be achieved by appropriate values of the applied field $B_0$, magnet size $R$, and relative distance $r_q$ between the NV and the magnet (see~\figref{Fig:Scaling}.c).
With this assumption, the coupling between the magnet and the higher level $\ket{1}$ is negligible, and the NV-magnet coupling is thus well approximated by $\Hop_{-1}$.
Hence, within the two-level approximation and the Holstein-Primakoff approximation the magnon-NV center Hamiltonian is given by
 \be\label{eq:Hnode_Gen}
\begin{split}
	\Hop_\text{MQ} =& \hbar\wsig(\theta)\frac{\sigz}{2}+\hbar\frac{\sqrt{F}}{2}\xi(\theta)\pare{e^{\im\vphi}\sigm+\hc}\\
	& -\hbar \cpare{\spare{g(\theta)e^{-\im2\vphi} \fdop +\W(\theta)  \fop}\sigm +\hc}\\
	&+\hbar \xi(\theta)\pare{e^{\im\vphi}\fop+\hc}\frac{\sigz}{2},
\end{split}
\ee
where $\sigz\equiv \ketbra{-1}{-1} - \ketbra{0}{0}$, $\sigp = \ketbra{-1}{0}$, $\sigm\equiv \ketbra{0}{-1}$, and we defined the frequencies
\bea
	\wsig(\theta) &\equiv & \Delta_\text{NV}- \gq B_0 - \frac{\hbar \gr\gq\mu_0}{4\pi r_q^3}F\pare{3\cos^2\theta-1},\qquad\label{eq:w_sig_theta}\\
	\xi(\theta) &\equiv& \frac{3\hbar\gq\gr\mu_0}{4\pi r_q^3}\sqrt{2F}\sin\theta\cos\theta,\label{eq:xi_theta}\\
	\W(\theta) &\equiv & \frac{\hbar \gr\gq \mu_0}{8\pi r_q^3}\sqrt{F}(3\sin^2\theta-2),\label{eq:g_theta}\\
	g(\theta) &\equiv & \frac{3\hbar\gr\gq\mu_0}{8\pi r_q^3}\sqrt{F}\sin^2\theta.\label{eq:W_theta}
\eea
The spin qubit in \eqnref{eq:Hnode_Gen} can be diagonalized in terms of dressed states $\ket{\pm}$. These are obtained from the uncoupled state $\ket{0},\ket{-1}$ by the unitary transformation matrix 
\be
	\U =
	\begin{pmatrix}
	e^{-\im\vphi/2}\cos(\Theta/2) & -e^{-\im\vphi/2}\sin(\Theta/2)\\
	e^{\im\vphi/2}\sin(\Theta/2) & e^{\im\vphi/2}\cos(\Theta/2)
	\end{pmatrix},	 
\ee
where the angle $\Theta\in[0,\pi]$ is defined as 
\be
	\Theta = \left\{ 
	\begin{array}{ll}
		\pi-\arctan\pare{\frac{\sqrt{F} \abs{\xi(\theta)}}{\abs{\wsig(\theta)}}},& \xi(\theta)/\wsig(\theta)<0\\
		\arctan\pare{\frac{\sqrt{F} \abs{\xi(\theta)}}{\abs{\wsig(\theta)}}},& \xi(\theta)/\wsig(\theta)>0
	\end{array}
	\right. .
\ee
In the dressed state basis, \eqnref{eq:Hnode_Gen} reads
\be\label{eq:Hnode_Gen_Final}
\begin{split}
	\Hop_\text{MQ} =& \hbar\wsig\frac{\sigz}{2}+\hbar\xi\pare{e^{\im\vphi}\fop+\hc}\frac{\sigz}{2}\\
	& -\hbar g\pare{e^{\im\vphi} \fdop\ketbra{+}{-} +\hc}\\
	&-\hbar \W\pare{e^{\im\vphi} \fdop\ketbra{-}{+} +\hc}.
\end{split}
\ee
The Pauli operator in \eqnref{eq:Hnode_Gen_Final} refer now to the dressed states, namely $\sigz \equiv \ketbra{+}{+}-\ketbra{-}{-}$. The dressed state frequency is
\be
	\wsig \equiv \sqrt{\wsig(\theta)^2 + F\xi(\theta)^2},
\ee
and the spin qubit-magnon couplings are
\bea
	\xi &\equiv &\! \xi(\theta) e^{\im\vphi}\! \cos\!\Theta +\pare{\W(\theta) e^{-\im\vphi}+g(\theta) e^{2\im\vphi}}\sin\!\Theta,\label{eq:xi_Gen}\\
	g &\equiv &\! \inv{4}\xi(\theta)\sin\!\Theta\!+\!g(\theta)\cos^2(\!\Theta/2\!)\!-\!\W(\theta) \sin^2(\!\Theta\!/2\!),\label{eq:g_Gen}
\eea
and
\be\label{eq:W_Gen}
	\W \equiv \inv{4}\xi(\theta)\sin\!\Theta+\W(\theta) \cos(\Theta/2)-g(\theta)\sin(\Theta/2).
\ee
\figref{Fig:Coupling_Theta} shows the dependence of the coupling $\xi$, $g$, and $\W$ in \eqnref{eq:xi_Gen}, \eqnref{eq:g_Gen}, and \eqnref{eq:W_Gen} as function of $\theta$.

The Hamiltonian describing the general scenario of quantum emitters (NV centers) locally coupled to the magnets in a HML by magnetic dipole-dipole interaction reads $\Hop_\text{T} = \Hop_\text{M} + \sum_{j=1}^N\Hop_\text{MQ}^j$, where $\Hop_M$ is the Hamiltonian of the HML and $Hop_\text{MQ}^j$ is the magnet-NV interaction at site $j$ and is given by \eqnref{eq:Hnode_Gen_Final}.
For a quasi resonant interaction, $\Delta \equiv \w_0-\wsig \ll \wsig,\w_0$, the total Hamiltonian of the system, within the rotating wave approximation, is given by the following Jaynes-Cumming-Hubbard Hamiltonian
\be\label{eq:JCH}
\begin{split}
	\Hop_\text{T} =& \hbar \sum_{\jj}\!\spare{\w_0 \fdop_\jj\fop_\jj + \wsig\frac{\sigz_\jj}{2} - g \pare{\fdop_\jj \sigm_\jj +\fop_\jj\sigp_\jj}\!}\\
  & + \hbar\mathcal{J}\sum_{\bold{i}\neq \jj}\fop_\bold{i}\fdop_\jj,
\end{split}
\ee
where $\jj=(j_x,j_y)$ and $\bold{i}=(i_x,i_y)$ are the index labeling the nodes of a general 2D HML. Note that \eqnref{eq:JCH} is the real space representation of \eqnref{eq:JCH_kspace} in the main text.

\section{Effective spin-spin interaction through a magnonic quantum bus}\label{apdx:SpinSpin_Interaction}

In the following, we derive \eqnref{eq:HQQ} describing the magnon-mediated interaction between the two qubits,  and we obtain the figure of merit presented in~\secref{sec:SWAP}, which estimates the efficacy of a SWAP gate operation performed by the magnonic quantum bus.

We consider two spin qubits locally coupled by magnetic dipole-dipole interactions to two magnetic particles coupled by a superconducting loop resonator (see \figref{Fig:Scaling}.a).
The dynamics of the quantum state of the total system is described by the following master equation
\be\label{eq:ME}
\begin{split}
	\pa{t}\rhop =& -\frac{\im}{\hbar}[\Hop_\text{T},\rhop]+\kp\sum_{j=1}^2\pare{\fop_j\rhop\fdop_j - \frac{1}{2}\{\fdop_j\fop_j,\rhop\}}\\
	&+\gamma \sum_{j=1}^2\pare{\sigz_j\rhop\sigz_j - \rhop},
\end{split}
\ee
where $\rhop$ represents the quantum state of the two qubits and the magnons at site $j=1,2$, $\kp$ is the magnon damping rate, $\gamma\equiv \pi/T_2^*$ the qubit dephasing rate, and $\Hop_\text{T}$ is defined in \eqnref{eq:JCH} for the simple case where $i,j=1,2$.
In terms of the modes $\fop_\pm \equiv (\fop_1\pm\fop_2)/\sqrt{2}$ the Hamiltonian $\Hop_\text{T}$ reads
\be
\begin{split}
	\Hop_\text{T} =& \hbar \w_+ \fdop_+\fop_+ +\hbar \w_- \fdop_-\fop_- +\hbar\wsig \pare{\frac{\sigz_1}{2}+\frac{\sigz_2}{2}}\\
	&-\frac{g}{\sqrt{2}}\Big[\fop_+ (\sigp_1+\sigp_2)+\fop_-(\sigp_1-\sigp_2)\\
	&+\hc\Big],
\end{split}
\ee
where $\w_\pm \equiv \w_0 \pm \mathcal{J}$. The dissipative term in \eqnref{eq:ME} maintains the same structure whereas the magnonic operators $\fop_j,\fdop_j$ ($j=1,2$) are replaced by the normal modes $\fdop_\pm,\fop_\pm$.

In the limit of a large detuning between the spin qubits and the magnons, it is possible to adiabatically eliminate the magnonic degrees of freedom and obtain an effective master equation describing the effective dynamics of the spin qubits.
Transforming the master equation describing the total system via the unitary operator
\be
\begin{split}
	\hat{U} \equiv & \exp\Big\{-\frac{g}{\sqrt{2}\Delta}\spare{\fdop_+(\sigm_1+\sigm_2)-\hc}\\
	&-\frac{g}{\sqrt{2}(\Delta-2\mathcal{J})}\spare{\fdop_-(\sigm_1-\sigm_2)-\hc}\Big\},
\end{split}
\ee
keeping terms up to second order in $g/\Delta,g/(\Delta-2\mathcal{J})\ll1$ and 
projecting the result on the vacuum subspace of the magnons Hilbert space, one obtains
\be\label{eq:ME_eff}
\begin{split} 
	\pa{t}\rhop_\text{eff} =& -\frac{\im}{\hbar}[\Hop_\text{QQ},\rhop_\text{eff}]+ \keff\sum_{j=1}^2\mathcal{D}^{jj}_{\sigm}[\rhop_\text{eff}]\\
	&+ \Weff\sum_{i\neq j=1}^2\mathcal{D}^{ij}_{\sigm}[\rhop_\text{eff}]\\
	&+\gamma \sum_{j=1}^2\pare{\sigz_j\rhop\sigz_j - \rhop},
\end{split}	
\ee
where
\be\label{eq:Ham_QQ}
\begin{split}
	\Hop_\text{QQ}\equiv\frac{\hbar}{2}\wtsig(\sigz_1+\sigz_2)-\hbar\geff(\sigp_1\sigm_2+\sigm_1\sigp_2),
\end{split}
\ee
and
\be
	\mathcal{D}^{ij}_{\sigm} [\rhop_\text{eff}] \equiv \sigm_i\rhop_\text{eff}\sigp_j -\frac{1}{2}\{\sigp_j\sigm_i,\rhop_\text{eff}\}.
\ee
Here, $\rhop_\text{eff}$ represents the effective state of the two qubits and we defined the effective frequencies and decay rates
\bea
	\wtsig &\equiv & \wsig - g^2\pare{\frac{1}{\Delta-2\mathcal{J}}+\frac{1}{\Delta}},\label{eq:wseff}\\
	\geff &\equiv & g^2\pare{\frac{1}{\Delta}-\frac{1}{\Delta-2\mathcal{J}}},\label{eq:geff}\\
	\keff &\equiv & \kp\, g^2\frac{\Delta^2+(\Delta-2\mathcal{J})^2}{\Delta^2(\Delta-2\mathcal{J})^2},\label{eq:keff}\\
	\Weff &\equiv & \kp\, g^2\frac{\Delta^2-(\Delta-2\mathcal{J})^2}{\Delta^2(\Delta-2\mathcal{J})^2}.\label{eq:Geff}
\eea
The Hamiltonian \eqnref{eq:Ham_QQ} can be used to implement a long-range qubit-qubit interaction through the magnonic quantum bus provided by a HML. 

The intrinsic qubit dephasing $\gamma$ as well as the bus-induced effective qubit damping $\keff$ described in \eqnref{eq:ME_eff}, affect the performance of coherent exchange of excitations between the qubits. In the following, we describe the impact of these noise sources and derive a figure of merit for the performance of the coherent qubit coupling. 

Let us consider a SWAP gate which transfers an excitation from the first to the second qubit through the interaction described by \eqnref{eq:Ham_QQ}.
The performance of the gate can be estimated in terms of the quantum state fidelity~\citep{Jozsa1994} $\mathcal{F}(t)=(\text{Tr}\sqrt{\bra{\psi_\text{t}}\rhop(t)\ket{\psi_\text{t}}})^2$ between the state of the system $\rhop(t)$ after the evolution governed by \eqnref{eq:ME} and the target state $\ket{\psi_\text{t}}\equiv \ket{01}\otimes\ket{\text{vac}}$, where $\ket{\text{vac}}$ is the vacuum of the magnon bus and $\ket{01}$ is the two qubits state where only the second (target) qubit is excited. We assume that the system is initially prepared in the pure state $\rhop(0)=\ketbra{\psi}{\psi}$ with $\ket{\psi} \equiv \ket{10}\otimes\ket{\text{vac}}$, where only the first qubit is excited.
The performance of the SWAP gate can then be estimated by maximizing $\mathcal{F}(t)$ over the total evolution time $t$, and calculating the fidelity error $\ve \equiv 1 - \text{max}_t \spare{\mathcal{F}(t)}$ in the presence of noise. A numerical optimization of $\ve$ over all the relevant parameters of the system $g,\mathcal{J},\Delta,\kp,$ and $\gamma$ yields a figure of merit for the performance of the gate.

An analytical expression for the scaling of the optimal error can be obtained in the dispersive regime $g/\Delta,g/(\Delta-2\mathcal{J})\ll1$.  
In the strong coupling limit $\geff \gg \keff,\gamma$, $\ve$ scales approximately linearly with the decoherence rates as $\ve \approx \alpha_{\gamma} \gamma/\geff + \alpha_\kp \keff/\geff$~\citep{Schuetz2017b}, where the coefficients $\alpha_\gamma$ and $\alpha_\kp$ are assumed to be approximately independent on the detuning $\Delta$. This assumption can be numerically checked simulating the error scaling for different values of $\Delta$ for the same values of $\keff/\geff,\gamma/\geff\ll1$. After substituting \eqnref{eq:geff} into the linear expansion for $\ve$ one obtains
\be\label{eq:MaxFid}
 	\ve = -\alpha_\gamma\frac{\gamma \Delta(\Delta -2\mathcal{J})}{2 g^2 \mathcal{J}}-\alpha_\kappa \frac{\kappa}{2}\frac{\Delta^2+(\Delta-2\mathcal{J})}{\mathcal{J}\Delta(\Delta-2\mathcal{J})}.
\ee
For the optimal values $\Delta^* = \mathcal{J}^* = \sqrt{2\alpha_\kp g^2 \kp T_2^*/(\pi\alpha_{\gamma})}$~\footnote{Different solutions for $\text{d}\ve/\text{d}\Delta =0$ are possible, however only $\Delta=\mathcal{J}$ satisfies the condition $g/\mathcal{J}\ll 1$ for the dispersive regime.}, \eqnref{eq:MaxFid} reads
\be
	\ve = \sqrt{\frac{\alpha_\kp\alpha_{\gamma}}{2C_0}},
\ee
where the cooperativity $C$ is defined as
\be
	C_0 = \frac{g^2}{\gamma \kp}.
\ee

\begin{figure}
	\includegraphics[width=\columnwidth]{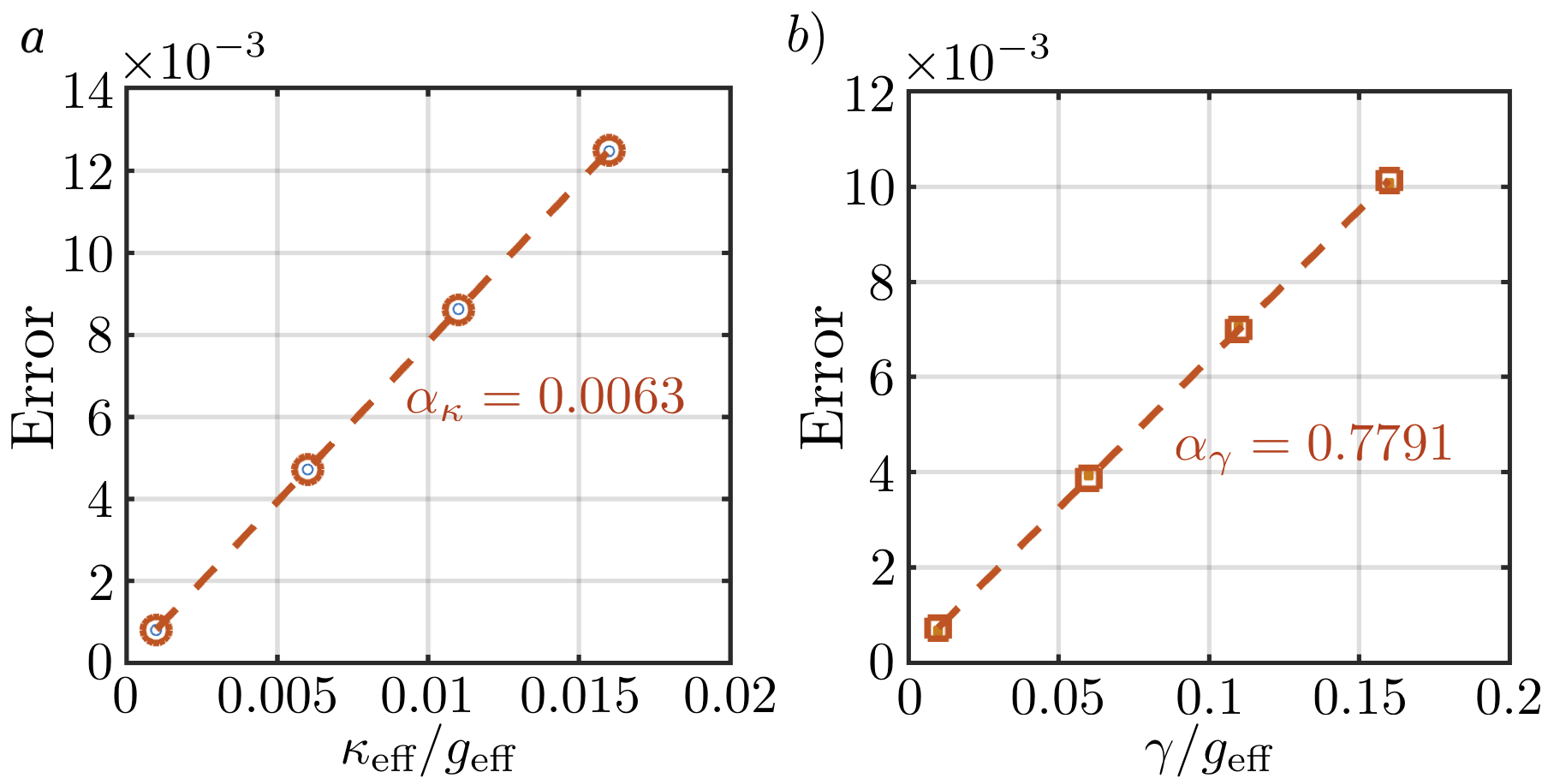}
	\caption{Distribution of the fidelity error $\ve$ for the optimized detuning $\Delta^*=\mathcal{J}$  as a function of a) $\kp/\geff$ for $\pi/T_2^*=0$ and b) $\gamma/\geff$ for for $\kp=0$. Each panel shows the numerically calculated points (square/circles) together with the linear interpolation (dashed line) with slope a) $\alpha_\kp$ and b) $\alpha_\gamma$.}
	\label{Fig:LinearReg} 
\end{figure}

In \figref{Fig:LinearReg}, the error $\ve$ optimized for $\Delta=\mathcal{J}$ is plotted as a function of the normalized decoherence rates $\keff/\geff$ and $\gamma/ \geff$.


\begin{thebibliography}{73}%
\makeatletter
\providecommand \@ifxundefined [1]{%
 \@ifx{#1\undefined}
}%
\providecommand \@ifnum [1]{%
 \ifnum #1\expandafter \@firstoftwo
 \else \expandafter \@secondoftwo
 \fi
}%
\providecommand \@ifx [1]{%
 \ifx #1\expandafter \@firstoftwo
 \else \expandafter \@secondoftwo
 \fi
}%
\providecommand \natexlab [1]{#1}%
\providecommand \enquote  [1]{``#1''}%
\providecommand \bibnamefont  [1]{#1}%
\providecommand \bibfnamefont [1]{#1}%
\providecommand \citenamefont [1]{#1}%
\providecommand \href@noop [0]{\@secondoftwo}%
\providecommand \href [0]{\begingroup \@sanitize@url \@href}%
\providecommand \@href[1]{\@@startlink{#1}\@@href}%
\providecommand \@@href[1]{\endgroup#1\@@endlink}%
\providecommand \@sanitize@url [0]{\catcode `\\12\catcode `\$12\catcode
  `\&12\catcode `\#12\catcode `\^12\catcode `\_12\catcode `\%12\relax}%
\providecommand \@@startlink[1]{}%
\providecommand \@@endlink[0]{}%
\providecommand \url  [0]{\begingroup\@sanitize@url \@url }%
\providecommand \@url [1]{\endgroup\@href {#1}{\urlprefix }}%
\providecommand \urlprefix  [0]{URL }%
\providecommand \Eprint [0]{\href }%
\providecommand \doibase [0]{http://dx.doi.org/}%
\providecommand \selectlanguage [0]{\@gobble}%
\providecommand \bibinfo  [0]{\@secondoftwo}%
\providecommand \bibfield  [0]{\@secondoftwo}%
\providecommand \translation [1]{[#1]}%
\providecommand \BibitemOpen [0]{}%
\providecommand \bibitemStop [0]{}%
\providecommand \bibitemNoStop [0]{.\EOS\space}%
\providecommand \EOS [0]{\spacefactor3000\relax}%
\providecommand \BibitemShut  [1]{\csname bibitem#1\endcsname}%
\let\auto@bib@innerbib\@empty
\bibitem [{\citenamefont {Kittel}(1987)}]{Kittel1987}%
  \BibitemOpen
  \bibfield  {author} {\bibinfo {author} {\bibfnamefont {C.}~\bibnamefont
  {Kittel}},\ }\href@noop {} {\emph {\bibinfo {title} {Quantum theory of
  solids}}}\ (\bibinfo  {publisher} {Wiley},\ \bibinfo {address} {New York},\
  \bibinfo {year} {1987})\BibitemShut {NoStop}%
\bibitem [{\citenamefont {Fetter}\ and\ \citenamefont
  {Walecka}(2003)}]{Fetter2012}%
  \BibitemOpen
  \bibfield  {author} {\bibinfo {author} {\bibfnamefont {A.~L.}\ \bibnamefont
  {Fetter}}\ and\ \bibinfo {author} {\bibfnamefont {J.~D.}\ \bibnamefont
  {Walecka}},\ }\href@noop {} {\emph {\bibinfo {title} {Quantum theory of
  many-particle systems}}}\ (\bibinfo  {publisher} {Dover Publications Inc.},\
  \bibinfo {address} {New York},\ \bibinfo {year} {2003})\BibitemShut {NoStop}%
\bibitem [{\citenamefont {Joannopoulos}\ \emph {et~al.}(2011)\citenamefont
  {Joannopoulos}, \citenamefont {Johnson}, \citenamefont {Winn},\ and\
  \citenamefont {Meade}}]{Joannopoulos2011}%
  \BibitemOpen
  \bibfield  {author} {\bibinfo {author} {\bibfnamefont {J.~D.}\ \bibnamefont
  {Joannopoulos}}, \bibinfo {author} {\bibfnamefont {S.~G.}\ \bibnamefont
  {Johnson}}, \bibinfo {author} {\bibfnamefont {J.~N.}\ \bibnamefont {Winn}}, \
  and\ \bibinfo {author} {\bibfnamefont {R.~D.}\ \bibnamefont {Meade}},\
  }\href@noop {} {\emph {\bibinfo {title} {Photonic crystals: molding the flow
  of light}}}\ (\bibinfo  {publisher} {Princeton university press},\ \bibinfo
  {year} {2011})\BibitemShut {NoStop}%
\bibitem [{\citenamefont {Chang}\ \emph {et~al.}(2018)\citenamefont {Chang},
  \citenamefont {Douglas}, \citenamefont {Gonz\'alez-Tudela}, \citenamefont
  {Hung},\ and\ \citenamefont {Kimble}}]{Chang2018}%
  \BibitemOpen
  \bibfield  {author} {\bibinfo {author} {\bibfnamefont {D.~E.}\ \bibnamefont
  {Chang}}, \bibinfo {author} {\bibfnamefont {J.~S.}\ \bibnamefont {Douglas}},
  \bibinfo {author} {\bibfnamefont {A.}~\bibnamefont {Gonz\'alez-Tudela}},
  \bibinfo {author} {\bibfnamefont {C.-L.}\ \bibnamefont {Hung}}, \ and\
  \bibinfo {author} {\bibfnamefont {H.~J.}\ \bibnamefont {Kimble}},\ }\href
  {\doibase 10.1103/RevModPhys.90.031002} {\bibfield  {journal} {\bibinfo
  {journal} {Rev. Mod. Phys.}\ }\textbf {\bibinfo {volume} {90}},\ \bibinfo
  {pages} {031002} (\bibinfo {year} {2018})}\BibitemShut {NoStop}%
\bibitem [{\citenamefont {Douglas}\ \emph {et~al.}(2015)\citenamefont
  {Douglas}, \citenamefont {Habibian}, \citenamefont {Hung}, \citenamefont
  {Gorshkov}, \citenamefont {Kimble},\ and\ \citenamefont
  {Chang}}]{Douglas2015}%
  \BibitemOpen
  \bibfield  {author} {\bibinfo {author} {\bibfnamefont {J.~S.}\ \bibnamefont
  {Douglas}}, \bibinfo {author} {\bibfnamefont {H.}~\bibnamefont {Habibian}},
  \bibinfo {author} {\bibfnamefont {C.~L.}\ \bibnamefont {Hung}}, \bibinfo
  {author} {\bibfnamefont {A.~V.}\ \bibnamefont {Gorshkov}}, \bibinfo {author}
  {\bibfnamefont {H.~J.}\ \bibnamefont {Kimble}}, \ and\ \bibinfo {author}
  {\bibfnamefont {D.~E.}\ \bibnamefont {Chang}},\ }\href
  {http://dx.doi.org/10.1038/nphoton.2015.57} {\bibfield  {journal} {\bibinfo
  {journal} {Nature Photonics}\ }\textbf {\bibinfo {volume} {9}},\ \bibinfo
  {pages} {326 EP } (\bibinfo {year} {2015})}\BibitemShut {NoStop}%
\bibitem [{\citenamefont {Hood}\ \emph {et~al.}(2016)\citenamefont {Hood},
  \citenamefont {Goban}, \citenamefont {Asenjo-Garcia}, \citenamefont {Lu},
  \citenamefont {Yu}, \citenamefont {Chang},\ and\ \citenamefont
  {Kimble}}]{Hood2016}%
  \BibitemOpen
  \bibfield  {author} {\bibinfo {author} {\bibfnamefont {J.~D.}\ \bibnamefont
  {Hood}}, \bibinfo {author} {\bibfnamefont {A.}~\bibnamefont {Goban}},
  \bibinfo {author} {\bibfnamefont {A.}~\bibnamefont {Asenjo-Garcia}}, \bibinfo
  {author} {\bibfnamefont {M.}~\bibnamefont {Lu}}, \bibinfo {author}
  {\bibfnamefont {S.-P.}\ \bibnamefont {Yu}}, \bibinfo {author} {\bibfnamefont
  {D.~E.}\ \bibnamefont {Chang}}, \ and\ \bibinfo {author} {\bibfnamefont
  {H.~J.}\ \bibnamefont {Kimble}},\ }\href
  {http://www.pnas.org/content/early/2016/08/30/1603788113} {\bibfield
  {journal} {\bibinfo  {journal} {Proceedings of the National Academy of
  Sciences}\ }\textbf {\bibinfo {volume} {113}},\ \bibinfo {pages} {10507}
  (\bibinfo {year} {2016})}\BibitemShut {NoStop}%
\bibitem [{\citenamefont {Hung}\ \emph {et~al.}(2016)\citenamefont {Hung},
  \citenamefont {Gonz{\'a}lez-Tudela}, \citenamefont {Cirac},\ and\
  \citenamefont {Kimble}}]{Hung2016}%
  \BibitemOpen
  \bibfield  {author} {\bibinfo {author} {\bibfnamefont {C.-L.}\ \bibnamefont
  {Hung}}, \bibinfo {author} {\bibfnamefont {A.}~\bibnamefont
  {Gonz{\'a}lez-Tudela}}, \bibinfo {author} {\bibfnamefont {J.~I.}\
  \bibnamefont {Cirac}}, \ and\ \bibinfo {author} {\bibfnamefont {H.~J.}\
  \bibnamefont {Kimble}},\ }\href {http://www.pnas.org/content/113/34/E4946}
  {\bibfield  {journal} {\bibinfo  {journal} {Proceedings of the National
  Academy of Sciences}\ }\textbf {\bibinfo {volume} {113}},\ \bibinfo {pages}
  {E4946} (\bibinfo {year} {2016})}\BibitemShut {NoStop}%
\bibitem [{\citenamefont {Schuetz}\ \emph
  {et~al.}(2017{\natexlab{a}})\citenamefont {Schuetz}, \citenamefont
  {Kn\"orzer}, \citenamefont {Giedke}, \citenamefont {Vandersypen},
  \citenamefont {Lukin},\ and\ \citenamefont {Cirac}}]{Schuetz2017}%
  \BibitemOpen
  \bibfield  {author} {\bibinfo {author} {\bibfnamefont {M.~J.~A.}\
  \bibnamefont {Schuetz}}, \bibinfo {author} {\bibfnamefont {J.}~\bibnamefont
  {Kn\"orzer}}, \bibinfo {author} {\bibfnamefont {G.}~\bibnamefont {Giedke}},
  \bibinfo {author} {\bibfnamefont {L.~M.~K.}\ \bibnamefont {Vandersypen}},
  \bibinfo {author} {\bibfnamefont {M.~D.}\ \bibnamefont {Lukin}}, \ and\
  \bibinfo {author} {\bibfnamefont {J.~I.}\ \bibnamefont {Cirac}},\ }\href
  {\doibase 10.1103/PhysRevX.7.041019} {\bibfield  {journal} {\bibinfo
  {journal} {Phys. Rev. X}\ }\textbf {\bibinfo {volume} {7}},\ \bibinfo {pages}
  {041019} (\bibinfo {year} {2017}{\natexlab{a}})}\BibitemShut {NoStop}%
\bibitem [{\citenamefont {Schuetz}\ \emph {et~al.}(2015)\citenamefont
  {Schuetz}, \citenamefont {Kessler}, \citenamefont {Giedke}, \citenamefont
  {Vandersypen}, \citenamefont {Lukin},\ and\ \citenamefont
  {Cirac}}]{Schuetz2015}%
  \BibitemOpen
  \bibfield  {author} {\bibinfo {author} {\bibfnamefont {M.~J.~A.}\
  \bibnamefont {Schuetz}}, \bibinfo {author} {\bibfnamefont {E.~M.}\
  \bibnamefont {Kessler}}, \bibinfo {author} {\bibfnamefont {G.}~\bibnamefont
  {Giedke}}, \bibinfo {author} {\bibfnamefont {L.~M.~K.}\ \bibnamefont
  {Vandersypen}}, \bibinfo {author} {\bibfnamefont {M.~D.}\ \bibnamefont
  {Lukin}}, \ and\ \bibinfo {author} {\bibfnamefont {J.~I.}\ \bibnamefont
  {Cirac}},\ }\href {\doibase 10.1103/PhysRevX.5.031031} {\bibfield  {journal}
  {\bibinfo  {journal} {Phys. Rev. X}\ }\textbf {\bibinfo {volume} {5}},\
  \bibinfo {pages} {031031} (\bibinfo {year} {2015})}\BibitemShut {NoStop}%
\bibitem [{\citenamefont {Yao}\ \emph {et~al.}(2012)\citenamefont {Yao},
  \citenamefont {Jiang}, \citenamefont {Gorshkov}, \citenamefont {Maurer},
  \citenamefont {Giedke}, \citenamefont {Cirac},\ and\ \citenamefont
  {Lukin}}]{Yao2012}%
  \BibitemOpen
  \bibfield  {author} {\bibinfo {author} {\bibfnamefont {N.~Y.}\ \bibnamefont
  {Yao}}, \bibinfo {author} {\bibfnamefont {L.}~\bibnamefont {Jiang}}, \bibinfo
  {author} {\bibfnamefont {A.~V.}\ \bibnamefont {Gorshkov}}, \bibinfo {author}
  {\bibfnamefont {P.~C.}\ \bibnamefont {Maurer}}, \bibinfo {author}
  {\bibfnamefont {G.}~\bibnamefont {Giedke}}, \bibinfo {author} {\bibfnamefont
  {J.~I.}\ \bibnamefont {Cirac}}, \ and\ \bibinfo {author} {\bibfnamefont
  {M.~D.}\ \bibnamefont {Lukin}},\ }\href
  {http://dx.doi.org/10.1038/ncomms1788} {\bibfield  {journal} {\bibinfo
  {journal} {Nature Communications}\ }\textbf {\bibinfo {volume} {3}},\
  \bibinfo {pages} {800 EP } (\bibinfo {year} {2012})}\BibitemShut {NoStop}%
\bibitem [{\citenamefont {Manenti}\ \emph {et~al.}(2017)\citenamefont
  {Manenti}, \citenamefont {Kockum}, \citenamefont {Patterson}, \citenamefont
  {Behrle}, \citenamefont {Rahamim}, \citenamefont {Tancredi}, \citenamefont
  {Nori},\ and\ \citenamefont {Leek}}]{Manenti2017}%
  \BibitemOpen
  \bibfield  {author} {\bibinfo {author} {\bibfnamefont {R.}~\bibnamefont
  {Manenti}}, \bibinfo {author} {\bibfnamefont {A.~F.}\ \bibnamefont {Kockum}},
  \bibinfo {author} {\bibfnamefont {A.}~\bibnamefont {Patterson}}, \bibinfo
  {author} {\bibfnamefont {T.}~\bibnamefont {Behrle}}, \bibinfo {author}
  {\bibfnamefont {J.}~\bibnamefont {Rahamim}}, \bibinfo {author} {\bibfnamefont
  {G.}~\bibnamefont {Tancredi}}, \bibinfo {author} {\bibfnamefont
  {F.}~\bibnamefont {Nori}}, \ and\ \bibinfo {author} {\bibfnamefont {P.~J.}\
  \bibnamefont {Leek}},\ }\href {\doibase 10.1038/s41467-017-01063-9}
  {\bibfield  {journal} {\bibinfo  {journal} {Nature Communications}\ }\textbf
  {\bibinfo {volume} {8}},\ \bibinfo {pages} {975} (\bibinfo {year}
  {2017})}\BibitemShut {NoStop}%
\bibitem [{\citenamefont {Gonz\'alez-Tudela}\ and\ \citenamefont
  {Cirac}(2017{\natexlab{a}})}]{GonzalezTudela2017PRA}%
  \BibitemOpen
  \bibfield  {author} {\bibinfo {author} {\bibfnamefont {A.}~\bibnamefont
  {Gonz\'alez-Tudela}}\ and\ \bibinfo {author} {\bibfnamefont {J.~I.}\
  \bibnamefont {Cirac}},\ }\href {\doibase 10.1103/PhysRevA.96.043811}
  {\bibfield  {journal} {\bibinfo  {journal} {Phys. Rev. A}\ }\textbf {\bibinfo
  {volume} {96}},\ \bibinfo {pages} {043811} (\bibinfo {year}
  {2017}{\natexlab{a}})}\BibitemShut {NoStop}%
\bibitem [{\citenamefont {John}\ and\ \citenamefont
  {Quang}(1994)}]{Sajeev1994}%
  \BibitemOpen
  \bibfield  {author} {\bibinfo {author} {\bibfnamefont {S.}~\bibnamefont
  {John}}\ and\ \bibinfo {author} {\bibfnamefont {T.}~\bibnamefont {Quang}},\
  }\href {\doibase 10.1103/PhysRevA.50.1764} {\bibfield  {journal} {\bibinfo
  {journal} {Phys. Rev. A}\ }\textbf {\bibinfo {volume} {50}},\ \bibinfo
  {pages} {1764} (\bibinfo {year} {1994})}\BibitemShut {NoStop}%
\bibitem [{\citenamefont {Huebl}\ \emph {et~al.}(2013)\citenamefont {Huebl},
  \citenamefont {Zollitsch}, \citenamefont {Lotze}, \citenamefont {Hocke},
  \citenamefont {Greifenstein}, \citenamefont {Marx}, \citenamefont {Gross},\
  and\ \citenamefont {Goennenwein}}]{Huebl2013}%
  \BibitemOpen
  \bibfield  {author} {\bibinfo {author} {\bibfnamefont {H.}~\bibnamefont
  {Huebl}}, \bibinfo {author} {\bibfnamefont {C.~W.}\ \bibnamefont
  {Zollitsch}}, \bibinfo {author} {\bibfnamefont {J.}~\bibnamefont {Lotze}},
  \bibinfo {author} {\bibfnamefont {F.}~\bibnamefont {Hocke}}, \bibinfo
  {author} {\bibfnamefont {M.}~\bibnamefont {Greifenstein}}, \bibinfo {author}
  {\bibfnamefont {A.}~\bibnamefont {Marx}}, \bibinfo {author} {\bibfnamefont
  {R.}~\bibnamefont {Gross}}, \ and\ \bibinfo {author} {\bibfnamefont
  {S.~T.~B.}\ \bibnamefont {Goennenwein}},\ }\href {\doibase
  10.1103/PhysRevLett.111.127003} {\bibfield  {journal} {\bibinfo  {journal}
  {Phys. Rev. Lett.}\ }\textbf {\bibinfo {volume} {111}},\ \bibinfo {pages}
  {127003} (\bibinfo {year} {2013})}\BibitemShut {NoStop}%
\bibitem [{\citenamefont {Tabuchi}\ \emph {et~al.}(2014)\citenamefont
  {Tabuchi}, \citenamefont {Ishino}, \citenamefont {Ishikawa}, \citenamefont
  {Yamazaki}, \citenamefont {Usami},\ and\ \citenamefont
  {Nakamura}}]{Tabuchi2014}%
  \BibitemOpen
  \bibfield  {author} {\bibinfo {author} {\bibfnamefont {Y.}~\bibnamefont
  {Tabuchi}}, \bibinfo {author} {\bibfnamefont {S.}~\bibnamefont {Ishino}},
  \bibinfo {author} {\bibfnamefont {T.}~\bibnamefont {Ishikawa}}, \bibinfo
  {author} {\bibfnamefont {R.}~\bibnamefont {Yamazaki}}, \bibinfo {author}
  {\bibfnamefont {K.}~\bibnamefont {Usami}}, \ and\ \bibinfo {author}
  {\bibfnamefont {Y.}~\bibnamefont {Nakamura}},\ }\href {\doibase
  10.1103/PhysRevLett.113.083603} {\bibfield  {journal} {\bibinfo  {journal}
  {Phys. Rev. Lett.}\ }\textbf {\bibinfo {volume} {113}},\ \bibinfo {pages}
  {083603} (\bibinfo {year} {2014})}\BibitemShut {NoStop}%
\bibitem [{\citenamefont {Lambert}\ \emph {et~al.}(2016)\citenamefont
  {Lambert}, \citenamefont {Haigh}, \citenamefont {Langenfeld}, \citenamefont
  {Doherty},\ and\ \citenamefont {Ferguson}}]{Lambert2016}%
  \BibitemOpen
  \bibfield  {author} {\bibinfo {author} {\bibfnamefont {N.~J.}\ \bibnamefont
  {Lambert}}, \bibinfo {author} {\bibfnamefont {J.~A.}\ \bibnamefont {Haigh}},
  \bibinfo {author} {\bibfnamefont {S.}~\bibnamefont {Langenfeld}}, \bibinfo
  {author} {\bibfnamefont {A.~C.}\ \bibnamefont {Doherty}}, \ and\ \bibinfo
  {author} {\bibfnamefont {A.~J.}\ \bibnamefont {Ferguson}},\ }\href {\doibase
  10.1103/PhysRevA.93.021803} {\bibfield  {journal} {\bibinfo  {journal} {Phys.
  Rev. A}\ }\textbf {\bibinfo {volume} {93}},\ \bibinfo {pages} {021803}
  (\bibinfo {year} {2016})}\BibitemShut {NoStop}%
\bibitem [{\citenamefont {Zhang}\ \emph {et~al.}(2016)\citenamefont {Zhang},
  \citenamefont {Zou}, \citenamefont {Jiang},\ and\ \citenamefont
  {Tang}}]{Zhang2016}%
  \BibitemOpen
  \bibfield  {author} {\bibinfo {author} {\bibfnamefont {X.}~\bibnamefont
  {Zhang}}, \bibinfo {author} {\bibfnamefont {C.}~\bibnamefont {Zou}}, \bibinfo
  {author} {\bibfnamefont {L.}~\bibnamefont {Jiang}}, \ and\ \bibinfo {author}
  {\bibfnamefont {H.~X.}\ \bibnamefont {Tang}},\ }\href {\doibase
  10.1063/1.4939134} {\bibfield  {journal} {\bibinfo  {journal} {Journal of
  Applied Physics}\ }\textbf {\bibinfo {volume} {119}},\ \bibinfo {pages}
  {023905} (\bibinfo {year} {2016})}\BibitemShut {NoStop}%
\bibitem [{\citenamefont {Zhang}\ \emph {et~al.}(2015)\citenamefont {Zhang},
  \citenamefont {Wang}, \citenamefont {Li}, \citenamefont {Luo}, \citenamefont
  {Wu}, \citenamefont {Nori},\ and\ \citenamefont {You}}]{Zhang2015}%
  \BibitemOpen
  \bibfield  {author} {\bibinfo {author} {\bibfnamefont {D.}~\bibnamefont
  {Zhang}}, \bibinfo {author} {\bibfnamefont {X.-M.}\ \bibnamefont {Wang}},
  \bibinfo {author} {\bibfnamefont {T.-F.}\ \bibnamefont {Li}}, \bibinfo
  {author} {\bibfnamefont {X.-Q.}\ \bibnamefont {Luo}}, \bibinfo {author}
  {\bibfnamefont {W.}~\bibnamefont {Wu}}, \bibinfo {author} {\bibfnamefont
  {F.}~\bibnamefont {Nori}}, \ and\ \bibinfo {author} {\bibfnamefont
  {J.}~\bibnamefont {You}},\ }\href {http://dx.doi.org/10.1038/npjqi.2015.14}
  {\bibfield  {journal} {\bibinfo  {journal} {Npj Quantum Information}\
  }\textbf {\bibinfo {volume} {1}},\ \bibinfo {pages} {15014 EP } (\bibinfo
  {year} {2015})}\BibitemShut {NoStop}%
\bibitem [{\citenamefont {Zare~Rameshti}\ \emph {et~al.}(2015)\citenamefont
  {Zare~Rameshti}, \citenamefont {Cao},\ and\ \citenamefont
  {Bauer}}]{Rameshti2015}%
  \BibitemOpen
  \bibfield  {author} {\bibinfo {author} {\bibfnamefont {B.}~\bibnamefont
  {Zare~Rameshti}}, \bibinfo {author} {\bibfnamefont {Y.}~\bibnamefont {Cao}},
  \ and\ \bibinfo {author} {\bibfnamefont {G.~E.~W.}\ \bibnamefont {Bauer}},\
  }\href {\doibase 10.1103/PhysRevB.91.214430} {\bibfield  {journal} {\bibinfo
  {journal} {Phys. Rev. B}\ }\textbf {\bibinfo {volume} {91}},\ \bibinfo
  {pages} {214430} (\bibinfo {year} {2015})}\BibitemShut {NoStop}%
\bibitem [{\citenamefont {Soykal}\ and\ \citenamefont
  {Flatt\'e}(2010)}]{Soykal2010}%
  \BibitemOpen
  \bibfield  {author} {\bibinfo {author} {\bibfnamefont {O.~O.}\ \bibnamefont
  {Soykal}}\ and\ \bibinfo {author} {\bibfnamefont {M.~E.}\ \bibnamefont
  {Flatt\'e}},\ }\href {\doibase 10.1103/PhysRevLett.104.077202} {\bibfield
  {journal} {\bibinfo  {journal} {Phys. Rev. Lett.}\ }\textbf {\bibinfo
  {volume} {104}},\ \bibinfo {pages} {077202} (\bibinfo {year}
  {2010})}\BibitemShut {NoStop}%
\bibitem [{\citenamefont {Viola~Kusminskiy}\ \emph {et~al.}(2016)\citenamefont
  {Viola~Kusminskiy}, \citenamefont {Tang},\ and\ \citenamefont
  {Marquardt}}]{Kusminskiy2016}%
  \BibitemOpen
  \bibfield  {author} {\bibinfo {author} {\bibfnamefont {S.}~\bibnamefont
  {Viola~Kusminskiy}}, \bibinfo {author} {\bibfnamefont {H.~X.}\ \bibnamefont
  {Tang}}, \ and\ \bibinfo {author} {\bibfnamefont {F.}~\bibnamefont
  {Marquardt}},\ }\href {\doibase 10.1103/PhysRevA.94.033821} {\bibfield
  {journal} {\bibinfo  {journal} {Phys. Rev. A}\ }\textbf {\bibinfo {volume}
  {94}},\ \bibinfo {pages} {033821} (\bibinfo {year} {2016})}\BibitemShut
  {NoStop}%
\bibitem [{\citenamefont {Osada}\ \emph {et~al.}(2016)\citenamefont {Osada},
  \citenamefont {Hisatomi}, \citenamefont {Noguchi}, \citenamefont {Tabuchi},
  \citenamefont {Yamazaki}, \citenamefont {Usami}, \citenamefont {Sadgrove},
  \citenamefont {Yalla}, \citenamefont {Nomura},\ and\ \citenamefont
  {Nakamura}}]{Osada2016}%
  \BibitemOpen
  \bibfield  {author} {\bibinfo {author} {\bibfnamefont {A.}~\bibnamefont
  {Osada}}, \bibinfo {author} {\bibfnamefont {R.}~\bibnamefont {Hisatomi}},
  \bibinfo {author} {\bibfnamefont {A.}~\bibnamefont {Noguchi}}, \bibinfo
  {author} {\bibfnamefont {Y.}~\bibnamefont {Tabuchi}}, \bibinfo {author}
  {\bibfnamefont {R.}~\bibnamefont {Yamazaki}}, \bibinfo {author}
  {\bibfnamefont {K.}~\bibnamefont {Usami}}, \bibinfo {author} {\bibfnamefont
  {M.}~\bibnamefont {Sadgrove}}, \bibinfo {author} {\bibfnamefont
  {R.}~\bibnamefont {Yalla}}, \bibinfo {author} {\bibfnamefont
  {M.}~\bibnamefont {Nomura}}, \ and\ \bibinfo {author} {\bibfnamefont
  {Y.}~\bibnamefont {Nakamura}},\ }\href {\doibase
  10.1103/PhysRevLett.116.223601} {\bibfield  {journal} {\bibinfo  {journal}
  {Phys. Rev. Lett.}\ }\textbf {\bibinfo {volume} {116}},\ \bibinfo {pages}
  {223601} (\bibinfo {year} {2016})}\BibitemShut {NoStop}%
\bibitem [{\citenamefont {Haigh}\ \emph {et~al.}(2018)\citenamefont {Haigh},
  \citenamefont {Lambert}, \citenamefont {Sharma}, \citenamefont {Blanter},
  \citenamefont {Bauer},\ and\ \citenamefont {Ramsay}}]{Haigh2018}%
  \BibitemOpen
  \bibfield  {author} {\bibinfo {author} {\bibfnamefont {J.~A.}\ \bibnamefont
  {Haigh}}, \bibinfo {author} {\bibfnamefont {N.~J.}\ \bibnamefont {Lambert}},
  \bibinfo {author} {\bibfnamefont {S.}~\bibnamefont {Sharma}}, \bibinfo
  {author} {\bibfnamefont {Y.~M.}\ \bibnamefont {Blanter}}, \bibinfo {author}
  {\bibfnamefont {G.~E.~W.}\ \bibnamefont {Bauer}}, \ and\ \bibinfo {author}
  {\bibfnamefont {A.~J.}\ \bibnamefont {Ramsay}},\ }\href {\doibase
  10.1103/PhysRevB.97.214423} {\bibfield  {journal} {\bibinfo  {journal} {Phys.
  Rev. B}\ }\textbf {\bibinfo {volume} {97}},\ \bibinfo {pages} {214423}
  (\bibinfo {year} {2018})}\BibitemShut {NoStop}%
\bibitem [{\citenamefont {Tabuchi}\ \emph {et~al.}(2015)\citenamefont
  {Tabuchi}, \citenamefont {Ishino}, \citenamefont {Noguchi}, \citenamefont
  {Ishikawa}, \citenamefont {Yamazaki}, \citenamefont {Usami},\ and\
  \citenamefont {Nakamura}}]{Tabuchi2015}%
  \BibitemOpen
  \bibfield  {author} {\bibinfo {author} {\bibfnamefont {Y.}~\bibnamefont
  {Tabuchi}}, \bibinfo {author} {\bibfnamefont {S.}~\bibnamefont {Ishino}},
  \bibinfo {author} {\bibfnamefont {A.}~\bibnamefont {Noguchi}}, \bibinfo
  {author} {\bibfnamefont {T.}~\bibnamefont {Ishikawa}}, \bibinfo {author}
  {\bibfnamefont {R.}~\bibnamefont {Yamazaki}}, \bibinfo {author}
  {\bibfnamefont {K.}~\bibnamefont {Usami}}, \ and\ \bibinfo {author}
  {\bibfnamefont {Y.}~\bibnamefont {Nakamura}},\ }\href {\doibase
  10.1126/science.aaa3693} {\bibfield  {journal} {\bibinfo  {journal}
  {Science}\ }\textbf {\bibinfo {volume} {349}},\ \bibinfo {pages} {405}
  (\bibinfo {year} {2015})}\BibitemShut {NoStop}%
\bibitem [{\citenamefont {Tabuchi}\ \emph {et~al.}(2016)\citenamefont
  {Tabuchi}, \citenamefont {Ishino}, \citenamefont {Noguchi}, \citenamefont
  {Ishikawa}, \citenamefont {Yamazaki}, \citenamefont {Usami},\ and\
  \citenamefont {Nakamura}}]{Tabuchi2016}%
  \BibitemOpen
  \bibfield  {author} {\bibinfo {author} {\bibfnamefont {Y.}~\bibnamefont
  {Tabuchi}}, \bibinfo {author} {\bibfnamefont {S.}~\bibnamefont {Ishino}},
  \bibinfo {author} {\bibfnamefont {A.}~\bibnamefont {Noguchi}}, \bibinfo
  {author} {\bibfnamefont {T.}~\bibnamefont {Ishikawa}}, \bibinfo {author}
  {\bibfnamefont {R.}~\bibnamefont {Yamazaki}}, \bibinfo {author}
  {\bibfnamefont {K.}~\bibnamefont {Usami}}, \ and\ \bibinfo {author}
  {\bibfnamefont {Y.}~\bibnamefont {Nakamura}},\ }\href {\doibase
  10.1016/j.crhy.2016.07.009} {\bibfield  {journal} {\bibinfo  {journal}
  {Comptes Rendus Physique}\ }\textbf {\bibinfo {volume} {17}},\ \bibinfo
  {pages} {729 } (\bibinfo {year} {2016})}\BibitemShut {NoStop}%
\bibitem [{\citenamefont {Nikitov}\ \emph {et~al.}(2001)\citenamefont
  {Nikitov}, \citenamefont {Tailhades},\ and\ \citenamefont
  {Tsai}}]{Nikitov2001}%
  \BibitemOpen
  \bibfield  {author} {\bibinfo {author} {\bibfnamefont {S.}~\bibnamefont
  {Nikitov}}, \bibinfo {author} {\bibfnamefont {P.}~\bibnamefont {Tailhades}},
  \ and\ \bibinfo {author} {\bibfnamefont {C.}~\bibnamefont {Tsai}},\ }\href
  {\doibase 10.1016/S0304-8853(01)00470-X} {\bibfield  {journal} {\bibinfo
  {journal} {Journal of Magnetism and Magnetic Materials}\ }\textbf {\bibinfo
  {volume} {236}},\ \bibinfo {pages} {320 } (\bibinfo {year}
  {2001})}\BibitemShut {NoStop}%
\bibitem [{\citenamefont {Krawczyk}\ and\ \citenamefont
  {Grundler}(2014)}]{Krawczyk2014}%
  \BibitemOpen
  \bibfield  {author} {\bibinfo {author} {\bibfnamefont {M.}~\bibnamefont
  {Krawczyk}}\ and\ \bibinfo {author} {\bibfnamefont {D.}~\bibnamefont
  {Grundler}},\ }\href {\doibase x10.1088/0953-8984/26/12/123202} {\bibfield
  {journal} {\bibinfo  {journal} {Journal of Physics: Condensed Matter}\
  }\textbf {\bibinfo {volume} {26}},\ \bibinfo {pages} {123202} (\bibinfo
  {year} {2014})}\BibitemShut {NoStop}%
\bibitem [{\citenamefont {Chumak}\ \emph {et~al.}(2017)\citenamefont {Chumak},
  \citenamefont {Serga},\ and\ \citenamefont {Hillebrands}}]{Chumak2017}%
  \BibitemOpen
  \bibfield  {author} {\bibinfo {author} {\bibfnamefont {A.}~\bibnamefont
  {Chumak}}, \bibinfo {author} {\bibfnamefont {A.}~\bibnamefont {Serga}}, \
  and\ \bibinfo {author} {\bibfnamefont {B.}~\bibnamefont {Hillebrands}},\
  }\href {\doibase 10.1088/1361-6463/aa6a65} {\bibfield  {journal} {\bibinfo
  {journal} {Journal of Physics D: Applied Physics}\ }\textbf {\bibinfo
  {volume} {50}},\ \bibinfo {pages} {244001} (\bibinfo {year}
  {2017})}\BibitemShut {NoStop}%
\bibitem [{\citenamefont {Andrich}\ \emph {et~al.}(2017)\citenamefont
  {Andrich}, \citenamefont {de~las Casas}, \citenamefont {Liu}, \citenamefont
  {Bretscher}, \citenamefont {Berman}, \citenamefont {Heremans}, \citenamefont
  {Nealey},\ and\ \citenamefont {Awschalom}}]{Andrich2017}%
  \BibitemOpen
  \bibfield  {author} {\bibinfo {author} {\bibfnamefont {P.}~\bibnamefont
  {Andrich}}, \bibinfo {author} {\bibfnamefont {C.~F.}\ \bibnamefont {de~las
  Casas}}, \bibinfo {author} {\bibfnamefont {X.}~\bibnamefont {Liu}}, \bibinfo
  {author} {\bibfnamefont {H.~L.}\ \bibnamefont {Bretscher}}, \bibinfo {author}
  {\bibfnamefont {J.~R.}\ \bibnamefont {Berman}}, \bibinfo {author}
  {\bibfnamefont {F.~J.}\ \bibnamefont {Heremans}}, \bibinfo {author}
  {\bibfnamefont {P.~F.}\ \bibnamefont {Nealey}}, \ and\ \bibinfo {author}
  {\bibfnamefont {D.~D.}\ \bibnamefont {Awschalom}},\ }\href {\doibase
  10.1038/s41534-017-0029-z} {\bibfield  {journal} {\bibinfo  {journal} {npj
  Quantum Information}\ }\textbf {\bibinfo {volume} {3}},\ \bibinfo {pages}
  {28} (\bibinfo {year} {2017})}\BibitemShut {NoStop}%
\bibitem [{\citenamefont {Kosen}\ \emph {et~al.}(2018)\citenamefont {Kosen},
  \citenamefont {Morris}, \citenamefont {van Loo},\ and\ \citenamefont
  {Karenowska}}]{Kosen2018}%
  \BibitemOpen
  \bibfield  {author} {\bibinfo {author} {\bibfnamefont {S.}~\bibnamefont
  {Kosen}}, \bibinfo {author} {\bibfnamefont {R.~G.~E.}\ \bibnamefont
  {Morris}}, \bibinfo {author} {\bibfnamefont {A.~F.}\ \bibnamefont {van Loo}},
  \ and\ \bibinfo {author} {\bibfnamefont {A.~D.}\ \bibnamefont {Karenowska}},\
  }\href {\doibase 10.1063/1.5011767} {\bibfield  {journal} {\bibinfo
  {journal} {Applied Physics Letters}\ }\textbf {\bibinfo {volume} {112}},\
  \bibinfo {pages} {012402} (\bibinfo {year} {2018})}\BibitemShut {NoStop}%
\bibitem [{\citenamefont {Trifunovic}\ \emph {et~al.}(2013)\citenamefont
  {Trifunovic}, \citenamefont {Pedrocchi},\ and\ \citenamefont
  {Loss}}]{Trifunovic2013}%
  \BibitemOpen
  \bibfield  {author} {\bibinfo {author} {\bibfnamefont {L.}~\bibnamefont
  {Trifunovic}}, \bibinfo {author} {\bibfnamefont {F.~L.}\ \bibnamefont
  {Pedrocchi}}, \ and\ \bibinfo {author} {\bibfnamefont {D.}~\bibnamefont
  {Loss}},\ }\href {\doibase 10.1103/PhysRevX.3.041023} {\bibfield  {journal}
  {\bibinfo  {journal} {Phys. Rev. X}\ }\textbf {\bibinfo {volume} {3}},\
  \bibinfo {pages} {041023} (\bibinfo {year} {2013})}\BibitemShut {NoStop}%
\bibitem [{\citenamefont {Qin}\ \emph {et~al.}(2018)\citenamefont {Qin},
  \citenamefont {Both}, \citenamefont {Hämäläinen}, \citenamefont {Yao},\
  and\ \citenamefont {van Dijken}}]{Qin2018}%
  \BibitemOpen
  \bibfield  {author} {\bibinfo {author} {\bibfnamefont {H.}~\bibnamefont
  {Qin}}, \bibinfo {author} {\bibfnamefont {G.-J.}\ \bibnamefont {Both}},
  \bibinfo {author} {\bibfnamefont {S.~J.}\ \bibnamefont {Hämäläinen}},
  \bibinfo {author} {\bibfnamefont {L.}~\bibnamefont {Yao}}, \ and\ \bibinfo
  {author} {\bibfnamefont {S.}~\bibnamefont {van Dijken}},\ }\href {\doibase
  10.1038/s41467-018-07893-5} {\bibfield  {journal} {\bibinfo  {journal}
  {Nature Communications}\ }\textbf {\bibinfo {volume} {9}},\ \bibinfo {pages}
  {5445} (\bibinfo {year} {2018})}\BibitemShut {NoStop}%
\bibitem [{\citenamefont {Topp}\ \emph {et~al.}(2010)\citenamefont {Topp},
  \citenamefont {Heitmann}, \citenamefont {Kostylev},\ and\ \citenamefont
  {Grundler}}]{Topp2010}%
  \BibitemOpen
  \bibfield  {author} {\bibinfo {author} {\bibfnamefont {J.}~\bibnamefont
  {Topp}}, \bibinfo {author} {\bibfnamefont {D.}~\bibnamefont {Heitmann}},
  \bibinfo {author} {\bibfnamefont {M.~P.}\ \bibnamefont {Kostylev}}, \ and\
  \bibinfo {author} {\bibfnamefont {D.}~\bibnamefont {Grundler}},\ }\href
  {\doibase 10.1103/PhysRevLett.104.207205} {\bibfield  {journal} {\bibinfo
  {journal} {Phys. Rev. Lett.}\ }\textbf {\bibinfo {volume} {104}},\ \bibinfo
  {pages} {207205} (\bibinfo {year} {2010})}\BibitemShut {NoStop}%
\bibitem [{\citenamefont {Gubbiotti}\ \emph {et~al.}(2007)\citenamefont
  {Gubbiotti}, \citenamefont {Tacchi}, \citenamefont {Carlotti}, \citenamefont
  {Singh}, \citenamefont {Goolaup}, \citenamefont {Adeyeye},\ and\
  \citenamefont {Kostylev}}]{Gubbiotti2007}%
  \BibitemOpen
  \bibfield  {author} {\bibinfo {author} {\bibfnamefont {G.}~\bibnamefont
  {Gubbiotti}}, \bibinfo {author} {\bibfnamefont {S.}~\bibnamefont {Tacchi}},
  \bibinfo {author} {\bibfnamefont {G.}~\bibnamefont {Carlotti}}, \bibinfo
  {author} {\bibfnamefont {N.}~\bibnamefont {Singh}}, \bibinfo {author}
  {\bibfnamefont {S.}~\bibnamefont {Goolaup}}, \bibinfo {author} {\bibfnamefont
  {A.~O.}\ \bibnamefont {Adeyeye}}, \ and\ \bibinfo {author} {\bibfnamefont
  {M.}~\bibnamefont {Kostylev}},\ }\href {\doibase 10.1063/1.2709909}
  {\bibfield  {journal} {\bibinfo  {journal} {Applied Physics Letters}\
  }\textbf {\bibinfo {volume} {90}},\ \bibinfo {pages} {092503} (\bibinfo
  {year} {2007})}\BibitemShut {NoStop}%
\bibitem [{\citenamefont {Davidovi\ifmmode~\acute{c}\else \'{c}\fi{}}\ \emph
  {et~al.}(1996)\citenamefont {Davidovi\ifmmode~\acute{c}\else \'{c}\fi{}},
  \citenamefont {Kumar}, \citenamefont {Reich}, \citenamefont {Siegel},
  \citenamefont {Field}, \citenamefont {Tiberio}, \citenamefont {Hey},\ and\
  \citenamefont {Ploog}}]{Davidovic1996}%
  \BibitemOpen
  \bibfield  {author} {\bibinfo {author} {\bibfnamefont {D.}~\bibnamefont
  {Davidovi\ifmmode~\acute{c}\else \'{c}\fi{}}}, \bibinfo {author}
  {\bibfnamefont {S.}~\bibnamefont {Kumar}}, \bibinfo {author} {\bibfnamefont
  {D.~H.}\ \bibnamefont {Reich}}, \bibinfo {author} {\bibfnamefont
  {J.}~\bibnamefont {Siegel}}, \bibinfo {author} {\bibfnamefont {S.~B.}\
  \bibnamefont {Field}}, \bibinfo {author} {\bibfnamefont {R.~C.}\ \bibnamefont
  {Tiberio}}, \bibinfo {author} {\bibfnamefont {R.}~\bibnamefont {Hey}}, \ and\
  \bibinfo {author} {\bibfnamefont {K.}~\bibnamefont {Ploog}},\ }\href
  {\doibase 10.1103/PhysRevLett.76.815} {\bibfield  {journal} {\bibinfo
  {journal} {Phys. Rev. Lett.}\ }\textbf {\bibinfo {volume} {76}},\ \bibinfo
  {pages} {815} (\bibinfo {year} {1996})}\BibitemShut {NoStop}%
\bibitem [{\citenamefont {Vool}\ and\ \citenamefont
  {Devoret}(2017)}]{Vool2017}%
  \BibitemOpen
  \bibfield  {author} {\bibinfo {author} {\bibfnamefont {U.}~\bibnamefont
  {Vool}}\ and\ \bibinfo {author} {\bibfnamefont {M.}~\bibnamefont {Devoret}},\
  }\href {\doibase 10.1002/cta.2359} {\bibfield  {journal} {\bibinfo  {journal}
  {International Journal of Circuit Theory and Applications}\ }\textbf
  {\bibinfo {volume} {45}},\ \bibinfo {pages} {897} (\bibinfo {year}
  {2017})}\BibitemShut {NoStop}%
\bibitem [{\citenamefont {De~Bernardis}\ \emph {et~al.}(2018)\citenamefont
  {De~Bernardis}, \citenamefont {Jaako},\ and\ \citenamefont
  {Rabl}}]{DeBernardis2018}%
  \BibitemOpen
  \bibfield  {author} {\bibinfo {author} {\bibfnamefont {D.}~\bibnamefont
  {De~Bernardis}}, \bibinfo {author} {\bibfnamefont {T.}~\bibnamefont {Jaako}},
  \ and\ \bibinfo {author} {\bibfnamefont {P.}~\bibnamefont {Rabl}},\ }\href
  {\doibase 10.1103/PhysRevA.97.043820} {\bibfield  {journal} {\bibinfo
  {journal} {Phys. Rev. A}\ }\textbf {\bibinfo {volume} {97}},\ \bibinfo
  {pages} {043820} (\bibinfo {year} {2018})}\BibitemShut {NoStop}%
\bibitem [{\citenamefont {Chikazumi}\ and\ \citenamefont
  {Graham}(2009)}]{Chikazumi}%
  \BibitemOpen
  \bibfield  {author} {\bibinfo {author} {\bibfnamefont {S.}~\bibnamefont
  {Chikazumi}}\ and\ \bibinfo {author} {\bibfnamefont {C.~D.}\ \bibnamefont
  {Graham}},\ }\href@noop {} {\emph {\bibinfo {title} {Physics of
  Ferromagnetism 2e}}}\ (\bibinfo  {publisher} {Oxford University Press on
  Demand},\ \bibinfo {address} {Oxford, UK},\ \bibinfo {year}
  {2009})\BibitemShut {NoStop}%
\bibitem [{Note1()}]{Note1}%
  \BibitemOpen
  \bibinfo {note} {We define the spin operator to be dimensionless, namely
  $[\protect \mathaccentV {hat}05E{F}_i,\protect \mathaccentV
  {hat}05E{F}_j]=\protect \text {i}\epsilon _{ijk}\protect \mathaccentV
  {hat}05E{F}_k$.}\BibitemShut {Stop}%
\bibitem [{Note2()}]{Note2}%
  \BibitemOpen
  \bibinfo {note} {With this configuration the field intensity at the surface
  of the coil generated by the magnet is $B \approx 50 \protect \text
  {mT}$~Appendix~\ref {apdx:critical_distance}. Depending on the distance
  between the particle and the loop might be necessary to consider the loop to
  be made of a high-Tc superconductor (see~Appendix~\ref
  {apdx:critical_distance}).}\BibitemShut {Stop}%
\bibitem [{\citenamefont {Gross}\ and\ \citenamefont
  {Haroche}(1982)}]{Gross1982}%
  \BibitemOpen
  \bibfield  {author} {\bibinfo {author} {\bibfnamefont {M.}~\bibnamefont
  {Gross}}\ and\ \bibinfo {author} {\bibfnamefont {S.}~\bibnamefont
  {Haroche}},\ }\href {\doibase 10.1016/0370-1573(82)90102-8} {\bibfield
  {journal} {\bibinfo  {journal} {Physics Reports}\ }\textbf {\bibinfo {volume}
  {93}},\ \bibinfo {pages} {301 } (\bibinfo {year} {1982})}\BibitemShut
  {NoStop}%
\bibitem [{\citenamefont {Grosso}\ and\ \citenamefont
  {Pastori~Parravicini}(2013)}]{Grosso2000}%
  \BibitemOpen
  \bibfield  {author} {\bibinfo {author} {\bibfnamefont {G.}~\bibnamefont
  {Grosso}}\ and\ \bibinfo {author} {\bibfnamefont {G.}~\bibnamefont
  {Pastori~Parravicini}},\ }\href@noop {} {\emph {\bibinfo {title} {Solid state
  physics}}}\ (\bibinfo  {publisher} {Elsevier Science Publishing Co Inc},\
  \bibinfo {address} {San Diego},\ \bibinfo {year} {2013})\BibitemShut
  {NoStop}%
\bibitem [{\citenamefont {Gonz\'alez-Tudela}\ and\ \citenamefont
  {Cirac}(2017{\natexlab{b}})}]{GonzalezTudela2017PRL}%
  \BibitemOpen
  \bibfield  {author} {\bibinfo {author} {\bibfnamefont {A.}~\bibnamefont
  {Gonz\'alez-Tudela}}\ and\ \bibinfo {author} {\bibfnamefont {J.~I.}\
  \bibnamefont {Cirac}},\ }\href {\doibase 10.1103/PhysRevLett.119.143602}
  {\bibfield  {journal} {\bibinfo  {journal} {Phys. Rev. Lett.}\ }\textbf
  {\bibinfo {volume} {119}},\ \bibinfo {pages} {143602} (\bibinfo {year}
  {2017}{\natexlab{b}})}\BibitemShut {NoStop}%
\bibitem [{\citenamefont {Gruber}\ \emph {et~al.}(1997)\citenamefont {Gruber},
  \citenamefont {Dr{\"a}benstedt}, \citenamefont {Tietz}, \citenamefont
  {Fleury}, \citenamefont {Wrachtrup},\ and\ \citenamefont
  {Borczyskowski}}]{Gruber2012}%
  \BibitemOpen
  \bibfield  {author} {\bibinfo {author} {\bibfnamefont {A.}~\bibnamefont
  {Gruber}}, \bibinfo {author} {\bibfnamefont {A.}~\bibnamefont
  {Dr{\"a}benstedt}}, \bibinfo {author} {\bibfnamefont {C.}~\bibnamefont
  {Tietz}}, \bibinfo {author} {\bibfnamefont {L.}~\bibnamefont {Fleury}},
  \bibinfo {author} {\bibfnamefont {J.}~\bibnamefont {Wrachtrup}}, \ and\
  \bibinfo {author} {\bibfnamefont {C.~v.}\ \bibnamefont {Borczyskowski}},\
  }\href {\doibase 10.1126/science.276.5321.2012} {\bibfield  {journal}
  {\bibinfo  {journal} {Science}\ }\textbf {\bibinfo {volume} {276}},\ \bibinfo
  {pages} {2012} (\bibinfo {year} {1997})}\BibitemShut {NoStop}%
\bibitem [{\citenamefont {Angelakis}\ \emph {et~al.}(2007)\citenamefont
  {Angelakis}, \citenamefont {Santos},\ and\ \citenamefont
  {Bose}}]{Angelakis2007}%
  \BibitemOpen
  \bibfield  {author} {\bibinfo {author} {\bibfnamefont {D.~G.}\ \bibnamefont
  {Angelakis}}, \bibinfo {author} {\bibfnamefont {M.~F.}\ \bibnamefont
  {Santos}}, \ and\ \bibinfo {author} {\bibfnamefont {S.}~\bibnamefont
  {Bose}},\ }\href {\doibase 10.1103/PhysRevA.76.031805} {\bibfield  {journal}
  {\bibinfo  {journal} {Phys. Rev. A}\ }\textbf {\bibinfo {volume} {76}},\
  \bibinfo {pages} {031805} (\bibinfo {year} {2007})}\BibitemShut {NoStop}%
\bibitem [{\citenamefont {Greentree}\ \emph {et~al.}(2006)\citenamefont
  {Greentree}, \citenamefont {Tahan}, \citenamefont {Cole},\ and\ \citenamefont
  {Hollenberg}}]{Greentree2006}%
  \BibitemOpen
  \bibfield  {author} {\bibinfo {author} {\bibfnamefont {A.~D.}\ \bibnamefont
  {Greentree}}, \bibinfo {author} {\bibfnamefont {C.}~\bibnamefont {Tahan}},
  \bibinfo {author} {\bibfnamefont {J.~H.}\ \bibnamefont {Cole}}, \ and\
  \bibinfo {author} {\bibfnamefont {L.~C.~L.}\ \bibnamefont {Hollenberg}},\
  }\href {http://dx.doi.org/10.1038/nphys466} {\bibfield  {journal} {\bibinfo
  {journal} {Nature Physics}\ }\textbf {\bibinfo {volume} {2}},\ \bibinfo
  {pages} {856 EP } (\bibinfo {year} {2006})}\BibitemShut {NoStop}%
\bibitem [{\citenamefont {Zhao}\ \emph {et~al.}(2012)\citenamefont {Zhao},
  \citenamefont {Honert}, \citenamefont {Schmid}, \citenamefont {Klas},
  \citenamefont {Isoya}, \citenamefont {Markham}, \citenamefont {Twitchen},
  \citenamefont {Jelezko}, \citenamefont {Liu}, \citenamefont {Fedder},\ and\
  \citenamefont {Wrachtrup}}]{Zhao2012}%
  \BibitemOpen
  \bibfield  {author} {\bibinfo {author} {\bibfnamefont {N.}~\bibnamefont
  {Zhao}}, \bibinfo {author} {\bibfnamefont {J.}~\bibnamefont {Honert}},
  \bibinfo {author} {\bibfnamefont {B.}~\bibnamefont {Schmid}}, \bibinfo
  {author} {\bibfnamefont {M.}~\bibnamefont {Klas}}, \bibinfo {author}
  {\bibfnamefont {J.}~\bibnamefont {Isoya}}, \bibinfo {author} {\bibfnamefont
  {M.}~\bibnamefont {Markham}}, \bibinfo {author} {\bibfnamefont
  {D.}~\bibnamefont {Twitchen}}, \bibinfo {author} {\bibfnamefont
  {F.}~\bibnamefont {Jelezko}}, \bibinfo {author} {\bibfnamefont {R.-B.}\
  \bibnamefont {Liu}}, \bibinfo {author} {\bibfnamefont {H.}~\bibnamefont
  {Fedder}}, \ and\ \bibinfo {author} {\bibfnamefont {J.}~\bibnamefont
  {Wrachtrup}},\ }\href {http://dx.doi.org/10.1038/nnano.2012.152} {\bibfield
  {journal} {\bibinfo  {journal} {Nature Nanotechnology}\ }\textbf {\bibinfo
  {volume} {7}},\ \bibinfo {pages} {657 EP } (\bibinfo {year}
  {2012})}\BibitemShut {NoStop}%
\bibitem [{\citenamefont {Abobeih}\ \emph {et~al.}(2018)\citenamefont
  {Abobeih}, \citenamefont {Cramer}, \citenamefont {Bakker}, \citenamefont
  {Kalb}, \citenamefont {Markham}, \citenamefont {Twitchen},\ and\
  \citenamefont {Taminiau}}]{Abobeih2018}%
  \BibitemOpen
  \bibfield  {author} {\bibinfo {author} {\bibfnamefont {M.~H.}\ \bibnamefont
  {Abobeih}}, \bibinfo {author} {\bibfnamefont {J.}~\bibnamefont {Cramer}},
  \bibinfo {author} {\bibfnamefont {M.~A.}\ \bibnamefont {Bakker}}, \bibinfo
  {author} {\bibfnamefont {N.}~\bibnamefont {Kalb}}, \bibinfo {author}
  {\bibfnamefont {M.}~\bibnamefont {Markham}}, \bibinfo {author} {\bibfnamefont
  {D.~J.}\ \bibnamefont {Twitchen}}, \ and\ \bibinfo {author} {\bibfnamefont
  {T.~H.}\ \bibnamefont {Taminiau}},\ }\href {\doibase
  10.1038/s41467-018-04916-z} {\bibfield  {journal} {\bibinfo  {journal}
  {Nature Communications}\ }\textbf {\bibinfo {volume} {9}},\ \bibinfo {pages}
  {2552} (\bibinfo {year} {2018})}\BibitemShut {NoStop}%
\bibitem [{\citenamefont {Boventer}\ \emph {et~al.}(2018)\citenamefont
  {Boventer}, \citenamefont {Pfirrmann}, \citenamefont {Krause}, \citenamefont
  {Sch\"on}, \citenamefont {Kl\"aui},\ and\ \citenamefont
  {Weides}}]{Boventer2018}%
  \BibitemOpen
  \bibfield  {author} {\bibinfo {author} {\bibfnamefont {I.}~\bibnamefont
  {Boventer}}, \bibinfo {author} {\bibfnamefont {M.}~\bibnamefont {Pfirrmann}},
  \bibinfo {author} {\bibfnamefont {J.}~\bibnamefont {Krause}}, \bibinfo
  {author} {\bibfnamefont {Y.}~\bibnamefont {Sch\"on}}, \bibinfo {author}
  {\bibfnamefont {M.}~\bibnamefont {Kl\"aui}}, \ and\ \bibinfo {author}
  {\bibfnamefont {M.}~\bibnamefont {Weides}},\ }\href {\doibase
  10.1103/PhysRevB.97.184420} {\bibfield  {journal} {\bibinfo  {journal} {Phys.
  Rev. B}\ }\textbf {\bibinfo {volume} {97}},\ \bibinfo {pages} {184420}
  (\bibinfo {year} {2018})}\BibitemShut {NoStop}%
\bibitem [{\citenamefont {Klingler}\ \emph {et~al.}(2017)\citenamefont
  {Klingler}, \citenamefont {Maier-Flaig}, \citenamefont {Dubs}, \citenamefont
  {Surzhenko}, \citenamefont {Gross}, \citenamefont {Huebl}, \citenamefont
  {Goennenwein},\ and\ \citenamefont {Weiler}}]{Klingler2017}%
  \BibitemOpen
  \bibfield  {author} {\bibinfo {author} {\bibfnamefont {S.}~\bibnamefont
  {Klingler}}, \bibinfo {author} {\bibfnamefont {H.}~\bibnamefont
  {Maier-Flaig}}, \bibinfo {author} {\bibfnamefont {C.}~\bibnamefont {Dubs}},
  \bibinfo {author} {\bibfnamefont {O.}~\bibnamefont {Surzhenko}}, \bibinfo
  {author} {\bibfnamefont {R.}~\bibnamefont {Gross}}, \bibinfo {author}
  {\bibfnamefont {H.}~\bibnamefont {Huebl}}, \bibinfo {author} {\bibfnamefont
  {S.~T.~B.}\ \bibnamefont {Goennenwein}}, \ and\ \bibinfo {author}
  {\bibfnamefont {M.}~\bibnamefont {Weiler}},\ }\href {\doibase
  10.1063/1.4977423} {\bibfield  {journal} {\bibinfo  {journal} {Applied
  Physics Letters}\ }\textbf {\bibinfo {volume} {110}},\ \bibinfo {pages}
  {092409} (\bibinfo {year} {2017})}\BibitemShut {NoStop}%
\bibitem [{\citenamefont {Stancil}\ and\ \citenamefont
  {Prabhakar}(2009)}]{Stancil}%
  \BibitemOpen
  \bibfield  {author} {\bibinfo {author} {\bibfnamefont {D.~D.}\ \bibnamefont
  {Stancil}}\ and\ \bibinfo {author} {\bibfnamefont {A.}~\bibnamefont
  {Prabhakar}},\ }\href@noop {} {\emph {\bibinfo {title} {Spin waves}}}\
  (\bibinfo  {publisher} {Springer},\ \bibinfo {address} {New York},\ \bibinfo
  {year} {2009})\BibitemShut {NoStop}%
\bibitem [{Note3()}]{Note3}%
  \BibitemOpen
  \bibinfo {note} {We remark that the effective description in Eq.~\protect
  \textup {\hbox {\mathsurround \z@ \protect \normalfont (\ignorespaces \ref
  {eq:HQQ}\unskip \@@italiccorr )}} breaks down when the qubit is resonant with
  the magnonic modes, \protect \textit {i.e.} for $\Delta \simeq 0, 2\protect
  \mathcal {J}$ (see also~Fig.~\ref {Fig:Scaling}.b). In this cases the
  dynamics of the system is correctly described by a master equation for the
  total magnons-qubit system (see~Appendix~\ref {apdx:SpinSpin_Interaction} for
  details).}\BibitemShut {Stop}%
\bibitem [{\citenamefont {Nisoli}\ \emph {et~al.}(2013)\citenamefont {Nisoli},
  \citenamefont {Moessner},\ and\ \citenamefont {Schiffer}}]{Nisoli2013}%
  \BibitemOpen
  \bibfield  {author} {\bibinfo {author} {\bibfnamefont {C.}~\bibnamefont
  {Nisoli}}, \bibinfo {author} {\bibfnamefont {R.}~\bibnamefont {Moessner}}, \
  and\ \bibinfo {author} {\bibfnamefont {P.}~\bibnamefont {Schiffer}},\ }\href
  {\doibase 10.1103/RevModPhys.85.1473} {\bibfield  {journal} {\bibinfo
  {journal} {Rev. Mod. Phys.}\ }\textbf {\bibinfo {volume} {85}},\ \bibinfo
  {pages} {1473} (\bibinfo {year} {2013})}\BibitemShut {NoStop}%
\bibitem [{\citenamefont {Heyderman}\ and\ \citenamefont
  {Stamps}(2013)}]{Heyderman2013}%
  \BibitemOpen
  \bibfield  {author} {\bibinfo {author} {\bibfnamefont {L.~J.}\ \bibnamefont
  {Heyderman}}\ and\ \bibinfo {author} {\bibfnamefont {R.~L.}\ \bibnamefont
  {Stamps}},\ }\href {\doibase 10.1088/0953-8984/25/36/363201} {\bibfield
  {journal} {\bibinfo  {journal} {Journal of Physics: Condensed Matter}\
  }\textbf {\bibinfo {volume} {25}},\ \bibinfo {pages} {363201} (\bibinfo
  {year} {2013})}\BibitemShut {NoStop}%
\bibitem [{\citenamefont {Ling}\ \emph {et~al.}(1996)\citenamefont {Ling},
  \citenamefont {Lezec}, \citenamefont {Higgins}, \citenamefont {Tsai},
  \citenamefont {Fujita}, \citenamefont {Numata}, \citenamefont {Nakamura},
  \citenamefont {Ochiai}, \citenamefont {Tang}, \citenamefont {Chaikin},\ and\
  \citenamefont {Bhattacharya}}]{Ling1996}%
  \BibitemOpen
  \bibfield  {author} {\bibinfo {author} {\bibfnamefont {X.~S.}\ \bibnamefont
  {Ling}}, \bibinfo {author} {\bibfnamefont {H.~J.}\ \bibnamefont {Lezec}},
  \bibinfo {author} {\bibfnamefont {M.~J.}\ \bibnamefont {Higgins}}, \bibinfo
  {author} {\bibfnamefont {J.~S.}\ \bibnamefont {Tsai}}, \bibinfo {author}
  {\bibfnamefont {J.}~\bibnamefont {Fujita}}, \bibinfo {author} {\bibfnamefont
  {H.}~\bibnamefont {Numata}}, \bibinfo {author} {\bibfnamefont
  {Y.}~\bibnamefont {Nakamura}}, \bibinfo {author} {\bibfnamefont
  {Y.}~\bibnamefont {Ochiai}}, \bibinfo {author} {\bibfnamefont
  {C.}~\bibnamefont {Tang}}, \bibinfo {author} {\bibfnamefont {P.~M.}\
  \bibnamefont {Chaikin}}, \ and\ \bibinfo {author} {\bibfnamefont
  {S.}~\bibnamefont {Bhattacharya}},\ }\href {\doibase
  10.1103/PhysRevLett.76.2989} {\bibfield  {journal} {\bibinfo  {journal}
  {Phys. Rev. Lett.}\ }\textbf {\bibinfo {volume} {76}},\ \bibinfo {pages}
  {2989} (\bibinfo {year} {1996})}\BibitemShut {NoStop}%
\bibitem [{\citenamefont {Berger}\ and\ \citenamefont
  {Rubinstein}(2001)}]{Berger2001}%
  \BibitemOpen
  \bibfield  {author} {\bibinfo {author} {\bibfnamefont {J.}~\bibnamefont
  {Berger}}\ and\ \bibinfo {author} {\bibfnamefont {J.}~\bibnamefont
  {Rubinstein}},\ }\href@noop {} {\emph {\bibinfo {title} {Connectivity and
  superconductivity}}}\ (\bibinfo  {publisher} {Springer Science \& Business
  Media},\ \bibinfo {address} {Berlin, Germany},\ \bibinfo {year}
  {2001})\BibitemShut {NoStop}%
\bibitem [{\citenamefont {Prat-Camps}\ \emph {et~al.}(2017)\citenamefont
  {Prat-Camps}, \citenamefont {Teo}, \citenamefont {Rusconi}, \citenamefont
  {Wieczorek},\ and\ \citenamefont {Romero-Isart}}]{PratCamps2017}%
  \BibitemOpen
  \bibfield  {author} {\bibinfo {author} {\bibfnamefont {J.}~\bibnamefont
  {Prat-Camps}}, \bibinfo {author} {\bibfnamefont {C.}~\bibnamefont {Teo}},
  \bibinfo {author} {\bibfnamefont {C.~C.}\ \bibnamefont {Rusconi}}, \bibinfo
  {author} {\bibfnamefont {W.}~\bibnamefont {Wieczorek}}, \ and\ \bibinfo
  {author} {\bibfnamefont {O.}~\bibnamefont {Romero-Isart}},\ }\href {\doibase
  10.1103/PhysRevApplied.8.034002} {\bibfield  {journal} {\bibinfo  {journal}
  {Phys. Rev. Applied}\ }\textbf {\bibinfo {volume} {8}},\ \bibinfo {pages}
  {034002} (\bibinfo {year} {2017})}\BibitemShut {NoStop}%
\bibitem [{\citenamefont {Rusconi}\ \emph {et~al.}(2017)\citenamefont
  {Rusconi}, \citenamefont {P\"ochhacker}, \citenamefont {Kustura},
  \citenamefont {Cirac},\ and\ \citenamefont {Romero-Isart}}]{Rusconi2017b}%
  \BibitemOpen
  \bibfield  {author} {\bibinfo {author} {\bibfnamefont {C.~C.}\ \bibnamefont
  {Rusconi}}, \bibinfo {author} {\bibfnamefont {V.}~\bibnamefont
  {P\"ochhacker}}, \bibinfo {author} {\bibfnamefont {K.}~\bibnamefont
  {Kustura}}, \bibinfo {author} {\bibfnamefont {J.~I.}\ \bibnamefont {Cirac}},
  \ and\ \bibinfo {author} {\bibfnamefont {O.}~\bibnamefont {Romero-Isart}},\
  }\href {\doibase 10.1103/PhysRevLett.119.167202} {\bibfield  {journal}
  {\bibinfo  {journal} {Phys. Rev. Lett.}\ }\textbf {\bibinfo {volume} {119}},\
  \bibinfo {pages} {167202} (\bibinfo {year} {2017})}\BibitemShut {NoStop}%
\bibitem [{\citenamefont {Goryachev}\ \emph {et~al.}(2014)\citenamefont
  {Goryachev}, \citenamefont {Farr}, \citenamefont {Creedon}, \citenamefont
  {Fan}, \citenamefont {Kostylev},\ and\ \citenamefont
  {Tobar}}]{Goryachev2014}%
  \BibitemOpen
  \bibfield  {author} {\bibinfo {author} {\bibfnamefont {M.}~\bibnamefont
  {Goryachev}}, \bibinfo {author} {\bibfnamefont {W.~G.}\ \bibnamefont {Farr}},
  \bibinfo {author} {\bibfnamefont {D.~L.}\ \bibnamefont {Creedon}}, \bibinfo
  {author} {\bibfnamefont {Y.}~\bibnamefont {Fan}}, \bibinfo {author}
  {\bibfnamefont {M.}~\bibnamefont {Kostylev}}, \ and\ \bibinfo {author}
  {\bibfnamefont {M.~E.}\ \bibnamefont {Tobar}},\ }\href {\doibase
  10.1103/PhysRevApplied.2.054002} {\bibfield  {journal} {\bibinfo  {journal}
  {Phys. Rev. Applied}\ }\textbf {\bibinfo {volume} {2}},\ \bibinfo {pages}
  {054002} (\bibinfo {year} {2014})}\BibitemShut {NoStop}%
\bibitem [{\citenamefont {Chang}\ and\ \citenamefont
  {Hsieh}(2004)}]{Chang2004}%
  \BibitemOpen
  \bibfield  {author} {\bibinfo {author} {\bibfnamefont {K.}~\bibnamefont
  {Chang}}\ and\ \bibinfo {author} {\bibfnamefont {L.-H.}\ \bibnamefont
  {Hsieh}},\ }\href@noop {} {\emph {\bibinfo {title} {Microwave ring circuits
  and related structures}}}\ (\bibinfo  {publisher} {John Wiley \& Sons},\
  \bibinfo {address} {New York},\ \bibinfo {year} {2004})\BibitemShut {NoStop}%
\bibitem [{\citenamefont {Hopkins}\ and\ \citenamefont
  {Free}(2008)}]{Hopkins2008}%
  \BibitemOpen
  \bibfield  {author} {\bibinfo {author} {\bibfnamefont {R.}~\bibnamefont
  {Hopkins}}\ and\ \bibinfo {author} {\bibfnamefont {C.}~\bibnamefont {Free}},\
  }\href@noop {} {\bibfield  {journal} {\bibinfo  {journal} {IET microwaves,
  antennas \& propagation}\ }\textbf {\bibinfo {volume} {2}},\ \bibinfo {pages}
  {66} (\bibinfo {year} {2008})}\BibitemShut {NoStop}%
\bibitem [{\citenamefont {Minev}\ \emph {et~al.}(2013)\citenamefont {Minev},
  \citenamefont {Pop},\ and\ \citenamefont {Devoret}}]{Minev2013}%
  \BibitemOpen
  \bibfield  {author} {\bibinfo {author} {\bibfnamefont {Z.~K.}\ \bibnamefont
  {Minev}}, \bibinfo {author} {\bibfnamefont {I.~M.}\ \bibnamefont {Pop}}, \
  and\ \bibinfo {author} {\bibfnamefont {M.~H.}\ \bibnamefont {Devoret}},\
  }\href {\doibase 10.1063/1.4824201} {\bibfield  {journal} {\bibinfo
  {journal} {Applied Physics Letters}\ }\textbf {\bibinfo {volume} {103}},\
  \bibinfo {pages} {142604} (\bibinfo {year} {2013})}\BibitemShut {NoStop}%
\bibitem [{\citenamefont {Minev}\ \emph {et~al.}(2016)\citenamefont {Minev},
  \citenamefont {Serniak}, \citenamefont {Pop}, \citenamefont {Leghtas},
  \citenamefont {Sliwa}, \citenamefont {Hatridge}, \citenamefont {Frunzio},
  \citenamefont {Schoelkopf},\ and\ \citenamefont {Devoret}}]{Minev2016}%
  \BibitemOpen
  \bibfield  {author} {\bibinfo {author} {\bibfnamefont {Z.~K.}\ \bibnamefont
  {Minev}}, \bibinfo {author} {\bibfnamefont {K.}~\bibnamefont {Serniak}},
  \bibinfo {author} {\bibfnamefont {I.~M.}\ \bibnamefont {Pop}}, \bibinfo
  {author} {\bibfnamefont {Z.}~\bibnamefont {Leghtas}}, \bibinfo {author}
  {\bibfnamefont {K.}~\bibnamefont {Sliwa}}, \bibinfo {author} {\bibfnamefont
  {M.}~\bibnamefont {Hatridge}}, \bibinfo {author} {\bibfnamefont
  {L.}~\bibnamefont {Frunzio}}, \bibinfo {author} {\bibfnamefont {R.~J.}\
  \bibnamefont {Schoelkopf}}, \ and\ \bibinfo {author} {\bibfnamefont {M.~H.}\
  \bibnamefont {Devoret}},\ }\href {\doibase 10.1103/PhysRevApplied.5.044021}
  {\bibfield  {journal} {\bibinfo  {journal} {Phys. Rev. Applied}\ }\textbf
  {\bibinfo {volume} {5}},\ \bibinfo {pages} {044021} (\bibinfo {year}
  {2016})}\BibitemShut {NoStop}%
\bibitem [{\citenamefont {Paik}\ \emph {et~al.}(2011)\citenamefont {Paik},
  \citenamefont {Schuster}, \citenamefont {Bishop}, \citenamefont {Kirchmair},
  \citenamefont {Catelani}, \citenamefont {Sears}, \citenamefont {Johnson},
  \citenamefont {Reagor}, \citenamefont {Frunzio}, \citenamefont {Glazman},
  \citenamefont {Girvin}, \citenamefont {Devoret},\ and\ \citenamefont
  {Schoelkopf}}]{Paik2011}%
  \BibitemOpen
  \bibfield  {author} {\bibinfo {author} {\bibfnamefont {H.}~\bibnamefont
  {Paik}}, \bibinfo {author} {\bibfnamefont {D.~I.}\ \bibnamefont {Schuster}},
  \bibinfo {author} {\bibfnamefont {L.~S.}\ \bibnamefont {Bishop}}, \bibinfo
  {author} {\bibfnamefont {G.}~\bibnamefont {Kirchmair}}, \bibinfo {author}
  {\bibfnamefont {G.}~\bibnamefont {Catelani}}, \bibinfo {author}
  {\bibfnamefont {A.~P.}\ \bibnamefont {Sears}}, \bibinfo {author}
  {\bibfnamefont {B.~R.}\ \bibnamefont {Johnson}}, \bibinfo {author}
  {\bibfnamefont {M.~J.}\ \bibnamefont {Reagor}}, \bibinfo {author}
  {\bibfnamefont {L.}~\bibnamefont {Frunzio}}, \bibinfo {author} {\bibfnamefont
  {L.~I.}\ \bibnamefont {Glazman}}, \bibinfo {author} {\bibfnamefont {S.~M.}\
  \bibnamefont {Girvin}}, \bibinfo {author} {\bibfnamefont {M.~H.}\
  \bibnamefont {Devoret}}, \ and\ \bibinfo {author} {\bibfnamefont {R.~J.}\
  \bibnamefont {Schoelkopf}},\ }\href {\doibase 10.1103/PhysRevLett.107.240501}
  {\bibfield  {journal} {\bibinfo  {journal} {Phys. Rev. Lett.}\ }\textbf
  {\bibinfo {volume} {107}},\ \bibinfo {pages} {240501} (\bibinfo {year}
  {2011})}\BibitemShut {NoStop}%
\bibitem [{\citenamefont {Grover}(2009)}]{Grover2004}%
  \BibitemOpen
  \bibfield  {author} {\bibinfo {author} {\bibfnamefont {F.~W.}\ \bibnamefont
  {Grover}},\ }\href@noop {} {\emph {\bibinfo {title} {Inductance calculations:
  working formulas and tables}}}\ (\bibinfo  {publisher} {Dover Publications
  Inc.},\ \bibinfo {address} {New York},\ \bibinfo {year} {2009})\BibitemShut
  {NoStop}%
\bibitem [{Note4()}]{Note4}%
  \BibitemOpen
  \bibinfo {note} {Without loss of generality, we assumed no flux trapped in
  the loop at the initial time.}\BibitemShut {Stop}%
\bibitem [{\citenamefont {Miltat}\ \emph {et~al.}(2002)\citenamefont {Miltat},
  \citenamefont {Albuquerque},\ and\ \citenamefont {Thiaville}}]{Miltat}%
  \BibitemOpen
  \bibfield  {author} {\bibinfo {author} {\bibfnamefont {J.}~\bibnamefont
  {Miltat}}, \bibinfo {author} {\bibfnamefont {G.}~\bibnamefont {Albuquerque}},
  \ and\ \bibinfo {author} {\bibfnamefont {A.}~\bibnamefont {Thiaville}},\ }in\
  \href@noop {} {\emph {\bibinfo {booktitle} {Spin Dynamics in Confined
  Magnetic Structures I}}}\ (\bibinfo  {publisher} {Springer},\ \bibinfo
  {address} {Berlin, Germany},\ \bibinfo {year} {2002})\ pp.\ \bibinfo {pages}
  {1--33}\BibitemShut {NoStop}%
\bibitem [{\citenamefont {Jackson}(1998)}]{Jackson}%
  \BibitemOpen
  \bibfield  {author} {\bibinfo {author} {\bibfnamefont {J.~D.}\ \bibnamefont
  {Jackson}},\ }\href@noop {} {\emph {\bibinfo {title} {Classical
  electrodynamics}}}\ (\bibinfo  {publisher} {John Wiley \& Sons},\ \bibinfo
  {address} {New York},\ \bibinfo {year} {1998})\BibitemShut {NoStop}%
\bibitem [{Note5()}]{Note5}%
  \BibitemOpen
  \bibinfo {note} {For the elementary two-magnet configuration in Fig.~\ref
  {Fig:Illustration}.b $\theta _{12}=0, \varphi _{ij}=\pi /2$ thus the first
  term in Eq.~\protect \textup {\hbox {\mathsurround \z@ \protect \normalfont
  (\ignorespaces \ref {eq:eta}\unskip \@@italiccorr )}} cancels.}\BibitemShut
  {Stop}%
\bibitem [{\citenamefont {Roberts}(1976)}]{Roberts1976}%
  \BibitemOpen
  \bibfield  {author} {\bibinfo {author} {\bibfnamefont {B.~W.}\ \bibnamefont
  {Roberts}},\ }\href@noop {} {\bibfield  {journal} {\bibinfo  {journal}
  {Journal of Physical and Chemical Reference Data}\ }\textbf {\bibinfo
  {volume} {5}},\ \bibinfo {pages} {581} (\bibinfo {year} {1976})}\BibitemShut
  {NoStop}%
\bibitem [{\citenamefont {Jozsa}(1994)}]{Jozsa1994}%
  \BibitemOpen
  \bibfield  {author} {\bibinfo {author} {\bibfnamefont {R.}~\bibnamefont
  {Jozsa}},\ }\href {\doibase 10.1080/09500349414552171} {\bibfield  {journal}
  {\bibinfo  {journal} {Journal of Modern Optics}\ }\textbf {\bibinfo {volume}
  {41}},\ \bibinfo {pages} {2315} (\bibinfo {year} {1994})}\BibitemShut
  {NoStop}%
\bibitem [{\citenamefont {Schuetz}\ \emph
  {et~al.}(2017{\natexlab{b}})\citenamefont {Schuetz}, \citenamefont {Giedke},
  \citenamefont {Vandersypen},\ and\ \citenamefont {Cirac}}]{Schuetz2017b}%
  \BibitemOpen
  \bibfield  {author} {\bibinfo {author} {\bibfnamefont {M.~J.~A.}\
  \bibnamefont {Schuetz}}, \bibinfo {author} {\bibfnamefont {G.}~\bibnamefont
  {Giedke}}, \bibinfo {author} {\bibfnamefont {L.~M.~K.}\ \bibnamefont
  {Vandersypen}}, \ and\ \bibinfo {author} {\bibfnamefont {J.~I.}\ \bibnamefont
  {Cirac}},\ }\href {\doibase 10.1103/PhysRevA.95.052335} {\bibfield  {journal}
  {\bibinfo  {journal} {Phys. Rev. A}\ }\textbf {\bibinfo {volume} {95}},\
  \bibinfo {pages} {052335} (\bibinfo {year} {2017}{\natexlab{b}})}\BibitemShut
  {NoStop}%
\bibitem [{Note6()}]{Note6}%
  \BibitemOpen
  \bibinfo {note} {Different solutions for $\protect \text {d}\varepsilon
  /\protect \text {d}\Delta =0$ are possible, however only $\Delta =\protect
  \mathcal {J}$ satisfies the condition $g/\protect \mathcal {J}\ll 1$ for the
  dispersive regime.}\BibitemShut {Stop}%
\end{thebibliography}
\end{document}